\documentclass[twocolumn]{aastex62}

\usepackage{amsmath}
\usepackage{graphics}

\DeclareMathOperator\erf{erf}
\usepackage{xcolor, fontawesome, microtype}
\definecolor{twitterblue}{RGB}{64,153,255}
\definecolor{linkcolor}{rgb}{0.1216,0.4667,0.7059}

\usepackage{ mathrsfs }

\received{October 28, 2019}
\revised{December 20, 2019}
\accepted{\today}
\submitjournal{ApJ}

\shorttitle{Assembly of a Fully Automated C5 Planet Candidate Catalog Using EDI-Vetter}

\shortauthors{Zink et al. (2020)}

\begin{document}

\title{Scaling \emph{K2}. II. Assembly of a Fully Automated C5 Planet Candidate Catalog Using EDI-Vetter}

\correspondingauthor{Jon Zink}
\email{jzink@astro.ucla.edu}

\author[0000-0003-1848-2063]{Jon K. Zink}
\affiliation{Department of Physics and Astronomy, University of California, Los Angeles, CA 90095}
\affiliation{Caltech/IPAC-NASA Exoplanet Science Institute, Pasadena, CA 91125}

\author[0000-0003-3702-0382]{Kevin K. Hardegree-Ullman}
\affiliation{Caltech/IPAC-NASA Exoplanet Science Institute, Pasadena, CA 91125}

\author[0000-0002-8035-4778]{Jessie L. Christiansen}
\affiliation{Caltech/IPAC-NASA Exoplanet Science Institute, Pasadena, CA 91125}

\author[0000-0001-8189-0233]{Courtney D. Dressing}
\affiliation{Department of Astronomy, University of California, Berkeley, CA 94720}

\author[0000-0002-1835-1891]{Ian J. M. Crossfield}
\affiliation{Department of Physics, and Kavli Institute for Astrophysics and Space Research, Massachusetts Institute of Technology, Cambridge, MA 02139}
\affiliation{Department of Physics and Astronomy, University of Kansas, Lawrence, KS 66045 }

\author[0000-0003-0967-2893]{Erik A. Petigura}
\affiliation{Department of Physics and Astronomy, University of California, Los Angeles, CA 90095}

\author[0000-0001-5347-7062]{Joshua E. Schlieder}
\affiliation{Exoplanets and Stellar Astrophysics Laboratory, Code 667, NASA Goddard Space Flight Center, Greenbelt, MD 20771}

\author[0000-0002-5741-3047]{David R. Ciardi}
\affiliation{Caltech/IPAC-NASA Exoplanet Science Institute, Pasadena, CA 91125}

\begin{abstract}

We present a uniform transiting exoplanet candidate list for Campaign 5 of the \emph{K2} mission. This catalog contains 75 planets with 7 multi-planet systems (5 double, 1 triple, and 1 quadruple planet system). Within the range of our search, we find 8 previously undetected candidates with the remaining 66 candidates overlapping 51\% of the Kruse et al. study that manually vet Campaign 5 candidates. In order to vet our potential transit signals, we introduce the Exoplanet Detection Identification Vetter ({\tt EDI-Vetter}), which is a fully automated program able to determine if a transit signal should be labeled as a false positive or a planet candidate. This automation allows us to create a statistically uniform catalog, ideal for planet occurrence rate measurements. When tested, the vetting software is able to ensure our sample is 94.2\% reliable against systematic false positives. Additionally, we inject artificial transits at the light-curve-level of the raw \emph{K2} data and find the maximum completeness of our pipeline is 70\% before vetting and 60\% after vetting. For convenience of future occurrence rate studies, we include measurements of stellar noise (CDPP) and the three-transit window function for each target. This study is part of a larger survey of the \emph{K2} data set and the methodology which will be applied to the entirety of the \emph{K2} data set.

\end{abstract}

\keywords{catalogs -- planetary systems -- stars: general, surveys}


\section{Introduction} 
\label{sec:intro}
The \emph{Kepler} spacecraft was launched in 2009 and has found over $4,000$ transiting exoplanet candidates. The original mission lasted 4 years collecting photometric data on more than $150,000$ stars in a 116 square degree patch of the sky \citep{koc10,bor16}. As intended, the fixed position and continuous (29.42 minute cadence) photometry provided the ideal conditions for transit detection and occurrence rate studies \citep[see e.g.,][]{you11,how12, don13, pet13b,dres13,dres15, bur15,muld15,muld18}. With the final announcement of Data Release 25 \citep[DR25;][]{tho18} and data from Gaia Data Release 2 \citep[DR2;][]{gai18}, additional studies have been conducted using improved stellar parameters \citep{bry19,zin19,zin19b,har19,hsu19}, providing a baseline measurement for galactic exoplanet occurrence.  

Upon the failure of two reaction wheels, the spacecraft was no longer able to remain focused on the same field for extended periods of time, thus concluding the original mission. However, the spacecraft was able to take photometric data despite the telescope's pointing issues. With the telescope now looking at different fields of the galaxy for shorter periods of the time ($<80$ days), the \emph{K2} mission was born \citep{how14, cle16}. The \emph{K2} mission continued for 18 complete campaigns (fields) along the ecliptic, before eventually running out of fuel. Each campaign delivered a unique glimpse at a different part of the ecliptic plane, providing the opportunity to consider how these differences in galactic latitude, metallicity, and stellar age \citep{riz18} may play a role in the occurrence of exoplanets. Performing occurrence rate measurements with \emph{K2} will also allow us to combine data from \emph{Kepler} to increase our sample size for global occurrence rate measurements.

Several groups have constructed catalogs of different \emph{K2} fields yielding over $800$ new exoplanet candidates \citep{van16,bar16,ada16, cro16,pop16,dre17,pet18,liv18,may18,ye18,zin19c,kru19}. Almost all of these catalogs have been focused on Campaigns 1-10, leaving a vast trove of un-examined data in Campaigns 11-18. Furthermore, vetting of the \emph{K2} transit signals has been almost entirely done by eye. Many of the pointing issues the spacecraft experienced created artificial dips that mimic transit signals. \emph{K2} data are considerably less well-behaved than the \emph{Kepler} data due to this pointing jitter, leading to many high signal-to-noise ratio (SNR) systematic false positives. Previous vetting software used for the \emph{Kepler} catalogs, which were not tailored to these systematics, could have difficulty isolating these false positives, but they are easily detected by human examination. This lack of automation and repeatability, thus far, makes \emph{K2} planet occurrence calculations difficult to perform.

One notable attempt to automate the search of \emph{K2} data was performed by \citet{kos19}, where previously detected candidates were vetted in a partly automated fashion (automated metrics prioritized candidates which could then be verified with human examination). However, this method still requires some human interaction, making measurements of completeness and reliability difficult to achieve.     

It is essential for any planet occurrence calculation that we know, and can account for, the biases to which the empirical sample are subjected. The automated pipeline for \emph{Kepler} was able to directly measure various metrics that can account for this lack of completeness \citep{mul15,cou16,tho18}. To test how well the pipeline is at recovering transit signals, previous studies have injected artificial transit signals into the light curves and measured the number of signals recovered by the automated pipeline \citep{chr13,chr15,chr17}. This gave a direct measure of pipeline detection efficiency. \citet{bur17b} provided a measure of the target window function (the probability that a signal will provide three transits within the window of data available) and the one-sigma depth function (which is a measure of the stellar noise given the transit duration). Both of these tools help calculate the probability of detecting a given transiting exoplanet. Without these metrics, strong assumptions about a catalog must be made in order to produce any measure of occurrence, and any conclusions should be taken with caution. Finally, to determine how pure the DR25 candidate sample was, reliability tests were performed by both inverting and scrambling the light curves and testing the pipeline for false positives \citep{tho18}. Using a fully automated pipeline is the only way to achieve an accurate measure of systematic reliability.

Using the \emph{Kepler} data set as a baseline detectable planet population, \citet{dot19} estimates \emph{K2} should yield $1317\pm261$ detectable planets. However, the \emph{Kepler} field may be intrinsically unique, distorting such estimates. With all the current \emph{K2} candidates detected through non-automated vetting processes, it remains difficult to estimate what fraction of the potential candidates have been found. In this paper we provide a fully automated \emph{K2} pipeline and candidate sample for Campaign 5. We select Campaign 5 for testing because this field has the largest known sample of \emph{K2} planet candidates (246 candidates, as of 2019 October 7), allowing us to tune the vetting metrics to maximize our planet yield. This paper presents a methodology which is in the process of being applied to all the \emph{K2} campaigns. In Section \ref{sec:data} we discuss how our pipeline takes the raw flux data and removes the noise from the spacecraft before searching and finding transit-like signals. In Section \ref{sec:vetter} we introduce our fully automated vetting software which has been optimized for \emph{K2} systematics. In Section \ref{sec:cdpp} we present our measure of stellar noise for each target. In Section \ref{sec:injections} we inject artificial transit-like signals into the light curves in order to measure the completeness of our pipeline. In Section \ref{sec:reliability} we present the results of our pipeline reliability test. In Section \ref{sec:window} the results of our window function measurements are presented. In Section \ref{sec:candidates} we present our list of fully vetted candidates. In Section \ref{sec:conclusions} we provide concluding remarks on our pipeline and candidate list. 


\section{Data Processing} 
\label{sec:data}
In the following section we will discuss how our pipeline takes the raw photometric data and converts it into usable transit signals. In brief, the target pixel files are downloaded from MAST (Campaign 5, Data Releases 31) and processed with the {\tt EVEREST} Python package \citep{lug16,lug18}. The light curves then underwent two Gaussian Process regressions and a harmonic signal removal. Finally, the fully detrended data are examined for transit signals using the {\tt TERRA} software \citep{pet13b}.   

\subsection{{\tt EVEREST}}
The \emph{K2} data set provides unique systematics that make the detection of exoplanets more difficult than that of the original \emph{Kepler} mission. The spacecraft thrusters and roll cause periodic fluctuations in the photometric data. In a preliminary study, we compared different \emph{K2} detrending software: {\tt K2SFF} \citep{van14}, {\tt K2PHOT} \citep{pet13b}, and {\tt EVEREST}\footnote{\href{https://rodluger.github.io/EVEREST/}{https://rodluger.github.io/EVEREST/}} \citep{lug16,lug18}. We found that all of these algorithms had different performance issues and strengths. The {\tt DAVE} vetting software \citep{kos19} considers several different detrended light curves simultaneously for each target, but such a task is computationally expensive when considering the desire to directly execute each detrending algorithm. We found that the {\tt EVEREST} software provided good minimization of signal RMS and the most user-friendly software for spacecraft fluctuation removal. In overview, {\tt EVEREST} uses the calibrated pixel-level data to produce time-series photometric light curves and corrects for systematic noise in those light curves by examining pixel-level correlations. All the \emph{K2} Campaigns have been processed with {\tt EVEREST} (however the calibrated pixel-level data files are currently undergoing reprocessed as part of the \emph{K2} Global Uniform Reprocessing Effort\footnote{\href{https://keplerscience.arc.nasa.gov/k2-uniform-global-reprocessing-underway.html}{https://keplerscience.arc.nasa.gov/k2-uniform-global-reprocessing-underway.html}}, warranting an updated processing with {\tt EVEREST}). Light curves with {\tt EVEREST} processing are publicly accessible\footnote{\href{https://archive.stsci.edu/hlsp/everest}{https://archive.stsci.edu/hlsp/everest}}. However, to ensure a uniform treatment, we undertook the task of reprocessing the Campaign 5 light curves with {\tt EVEREST}, using the {\tt K2SFF} aperture \# 15, as suggested by \citet{lug18}. This aperture uses a model of the \emph{Kepler} Pixel Response Function to determine the size and shape of the target aperture \citep{van14, bry10}.  This gave us the ability to later inject signals into the raw flux data (light-curve-level) before {\tt EVEREST} processing (see Section \ref{sec:injections}). By directly running {\tt EVEREST} on the calibrated pixel-level data (downloaded from MAST)\footnote{\href{https://archive.stsci.edu/k2/}{https://archive.stsci.edu/k2/}}, we could ensure the transit injection recovery is processed using the exact same conditions as that of the candidate sample.

On average, {\tt EVEREST} masks about 8\% of the light curve due to spacecraft anomalies or outliers. Occasionally, {\tt EVEREST} will remove a large fraction of the data points within a light curve as the excessive noise is seen as numerous outliers. We remove targets where more that 50\% of the light curve is masked. This severe masking only occurs in 37 of the Campaign 5 light curves, in most cases where the target is near the edge of the CCD. These 37 targets represent a very small fraction of the 25,040 light curves considered for this campaign.

\subsection{GP and Harmonic Detrending}
\label{sec:har}

\begin{figure}
\centering \includegraphics[height=6.7cm]{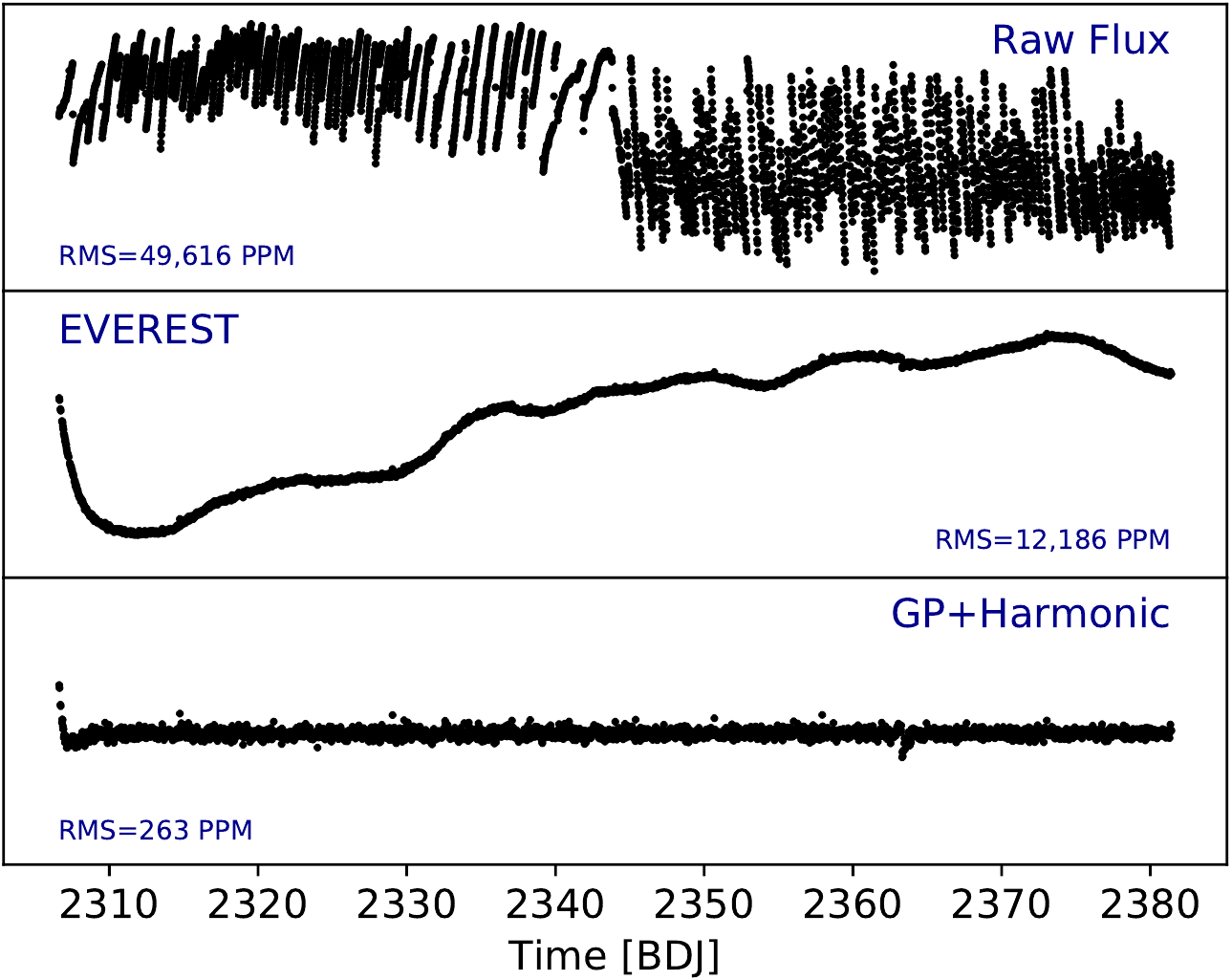}

\caption{\textbf{Top} The raw target pixel file data for EPIC 211422469. \textbf{Middle} shows the data after being processed with the {\tt EVEREST} detrending software. \textbf{Bottom} the light curve after an additional processing from our GP and harmonic fitter routine. The corresponding RMS values are shown, in units of parts per million (PPM) to illustrate how each step will condense the noise of the light curve.     \label{fig:process}}
\end{figure}

The {\tt EVEREST} software removes the systematics caused by the thrusters and roll of the spacecraft, but an additional global detrending is needed to flatten the light curves. We achieve this through Gaussian Process (GP) regression and harmonic removal.

The first step in detrending is to establish the white noise level ($\sigma$). It is important that the detrending knows what is signal versus noise, and an accurate $\sigma$ value will help achieve this goal. In well-behaved flat light curves this is an easy task, and $\sigma$ is the standard deviation of the normalized flux. However, if the time series contains correlated noise, the standard method will artificially increase the measured uncertainty. To complicate this even further, deep transit signals or a large number of outliers can again inflate the RMS. We use the median absolute deviation (MAD), which is a measure of the median absolute value distance from the median value in the time series. This value can be scaled to estimate $\sigma$ for the time series: 
\begin{equation}
\sigma_{\text{MAD}}=\frac{\text{MAD}}{\Phi^{-1}(3/4)}\approx 1.4826 \text{ MAD}
\label{eq:MAD}
\end{equation}
where $\Phi^{-1}(3/4)$ is the normal inverse cumulative distribution function evaluated at probability of 3/4. This measure of $\sigma$ is far more robust to outliers and transit signals. However, light curves that require significant red-noise removal will still inflate the inferred $\sigma$ value. To limit this issue we break the light curve into bins of width 1 day, as the daily flux trend should be minimal. We measure the $\sigma_{\text{MAD}}$ value for each of these bins and take the median $\sigma_{\text{MAD}}$ value from these bins as our measured value for $\sigma$. It is important to note that $<1$ days stellar variability can still inflate our measured $\sigma$ value, but this effect is insignificant when looking for long term trends as done here.

One challenge of GP detrending is that GPs can fit out real transit signals. A thorough discussion of this issue can be found in \citet{hip19}, with several alternative detrending methods which we hope to consider in future iterations of our pipeline. To avoid this issue, we attempt to mask out outliers and potential transits. We again bin the time series by single day intervals. Within each bin we look for points that exceed $3\sigma_{\text{MAD}}$ from the median of each bin. These points are only masked for the detrending process and unmasked for the rest of our pipeline. By manually adjusting the threshold, we found that a $3\sigma$ clipping was sufficient in masking most significant transits. Additionally, any transit existing below this limit was likely to be seen as noise by the GP, and therefore not removed. An additional restriction we place on our GP is that we do not allow it to choose detrending periods $<5$ days. Periods below this range are more prone to over-fitting transit signals as the transit duration is often a significant fraction of a day. Putting this limitation in place decreases our chances of modifying the transit depth. We discuss this issue more thoroughly in Section \ref{sec:candidates}. 

To implement our GP regression the pipeline uses {\tt PyMC3} \citep{sal15} with a wrapper built by the {\tt Exoplanet} software \citep{for19}. Our detrending uses the ``Rotation'' kernel which is a combination of two simple harmonic oscillators, meant to mimic stellar rotation/noise. Our pipeline uses this GP detrending in two steps: First, it looks for long term trends (periods $>10$ days and general flux drifting), then after removing the long term trends, it looks for short term trends ($5$ days $<$ period $<10$ days) and subtract away the best fit GP. However, stellar oscillations can also exist on very short timescales (often $<0.5$ days). Any attempt to remove these oscillations with a GP will almost certainly remove transits as well, thus a harmonic fitter is used to address this issue.

Unlike a GP, a harmonic fitter does not have the flexibility to fit and remove long period transits. It can only affect the signals with strong similarities to a sine function and the long periods of flat base line between transits makes this unlikely. This can become problematic when you have a system with multiple transiting planets, decreasing the gaps between transits and mimicking a sine curve \citep{chr13,zin19}. Additionally, short period giant planets can also be affected by this harmonic removal process, by either decreasing the transit depth or removing the signal completely. By choosing an upper limit of 0.5 days for our harmonic fitter, we minimize such occurrences. We begin by unmasking the GP sigma-clipped data and search for the strongest periodic signal below 0.5 days using a Lomb-Scargle periodogram. We then use a $\chi^2$ fitting algorithm to measure the amplitude ($A$) of the signal. It is important that we do not introduce additional noise into the data set, so we require $A>10\sigma_A$ before removing the harmonic signal. We found in practise that this limit rarely over-fit the data, but we address such possibilities in our vetting metrics (Section \ref{sec:Haromic}). In Figure \ref{fig:process} we show how the raw flux data will change at each step of our light curve processing.

Looking at our GP and harmonic fitter, there is an apparent gap in our considered period range for detrending ($0.5<$ periods $<5$ days). Any attempt made to fit out this period range either introduced additional artificial signals into our data, or artificially reduced transit depths. Fortunately, most stars have rotational periods $>5$ days \citep{mcq13} and most variable stars oscillate at periods $<0.5$ days, indicating that few non-planetary periodic signals occupy this parameter space. Additionally, our vetting algorithm rejects any signals found with such harmonic features, minimizing the possibility of a False Positive (FP) detection (Section \ref{sec:Haromic}).

\subsection{{\tt TERRA}}
\label{sec:terra}
Once the light curve has been fully processed, we begin our search for Threshold Crossing Events (TCEs). These are signals with at least three transit-like events that reach or exceed a specific statistical significance. We discuss our process of selecting this threshold in the next Section (\ref{sec:tceLimit}). To search for these signals, we use the {\tt TERRA} search algorithm \citep{pet13b}. This software begins by masking outliers and then uses a finely-spaced grid-based search algorithm to find the largest signal-to-noise ratio (SNR) event in phase-folded period space of the light curve. This 3D grid search provides a measure of period ($P$), the transit duration ($t_{dur}$), the transit emphemeris ($t_0$), and SNR. For transiting planets detected by the \emph{Kepler} pipeline, the SNR is measured using the Multiple Event Statistic \citep[MES;][]{je02}, which indicates the strength of the whitened signal assuming a linear emphemeris. We will use MES in reference to the strength of the signal henceforth.\footnote{We note that the SES and MES used by {\tt TERRA} and \emph{Kepler} project Transiting Planet Search (TPS) are analogous, but not strictly equivalent. Differences in the detailed constructions of these statistics may be found by comparing \citet{jen10} and \citet{pet13b}.}

To recover systems with more than one transiting exoplanet, we consider the five signals with the largest MES in the light curve. We do not expect many systems to contain more than five distinct detectable transits within a period of 38 days (\emph{Kepler}-80 and \emph{TRAPPIST}-1 are currently the only known exceptions). However, \citet{kru19} found evidence for one such system (EPIC 210965800) with six planets in Campaign 4 of \emph{K2}. We acknowledge that we may lose some of these higher multiplicity systems by limiting our search to five planets, but such a restriction is necessary to minimize our computational cost. After the largest MES signal has been detected, we then mask $2.5\times$ the transit duration ($1.25t_{dur}$ on either side of predicted transit midpoint), allowing the next largest signal to be detected. This process continues until either five TCEs have been found, or the largest MES value is less than the detection threshold. \citet{zin19} showed that this type of masking will make higher multiplicity systems more difficult to detect, as nearby transits have the potential of being partially or fully hidden by this type of masking. However, removing the signal with a transit model is impractical. A majority of the signals are FPs that do not match well with the transit model. Thus, subtracting the best fit model will not actually remove the signal, but rather morph the data so that the grid search continues to detect the same anomaly repeatedly. In addition, any poorly-fit real transit signal will not be completely removed through model subtraction, making smaller MES signals very difficult to find. Therefore, masking is the only way to ensure the previous signal is not being re-detected at each iteration of multiplicity ($m$). To minimize the effects of masking we choose to only remove $2.5\times$ the transit duration, whereas the \emph{Kepler} pipeline used a mask of $3\times$ the transit duration, which should decrease the probability of masking neighboring transits.

Searching for transits, we limit the orbital period search range to [0.5, 38] days. The upper limit of 38 days is set by the span of the data (nominally $74.84$ days for Campaign 5), which permits the existence of three transits within the given window. The lower limit was set to minimize contamination from the abundance of harmonic signals that exist at periods $<0.5$ days ({\tt TERRA} can detect signals with sufficient MES and periods $<0.5$ days at a multiple of the true period, but in practise we found these are more often FPs than true planets). Additionally, we wanted to minimize the probability of finding signals artificially introduced by the harmonic fitter (Section \ref{sec:har}). Finally, of the over 4000 confirmed exoplanets, only 15 have orbital periods $<0.5$ days (0.37\%), including six planets found in the \emph{K2} fields. Thus, limiting our search should have a small effect on the extracted period population. Users concerned with the occurrence of ultra-short period (USP) planets are encouraged to perform their own custom search.

\subsection{Selecting a TCE Limit}
\label{sec:tceLimit}

\begin{figure}
\centering \includegraphics[height=6.7cm]{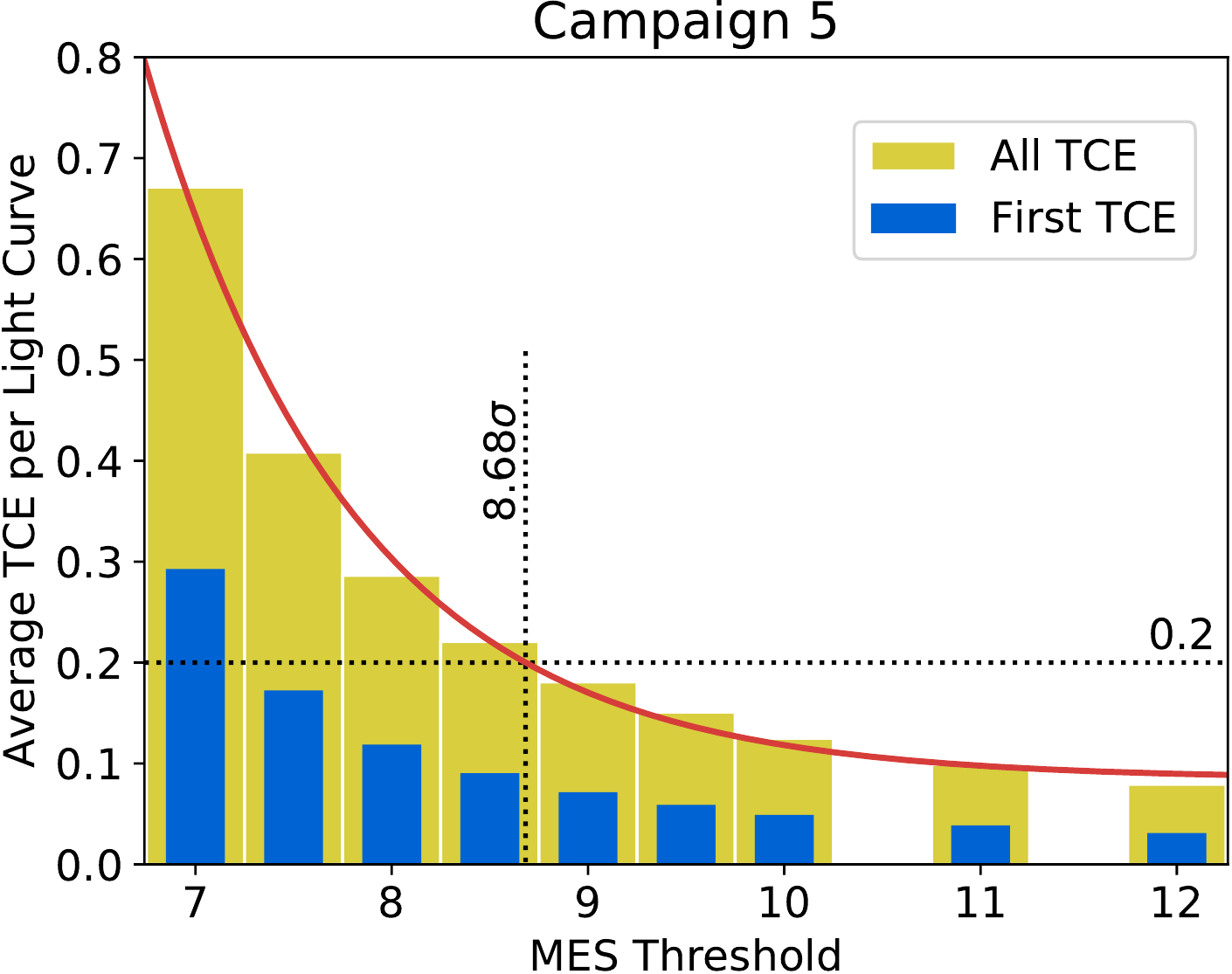}

\caption{MES threshold plotted against the average number of TCEs found for each light curve in Campaign 5. Many of the light curves will produce more than one TCE, thus the yellow bars (All TCE) represent the total number of detected TCEs divided by the number of light curves searched. The blue bars (First TCE) only consider the first TCE found in each light curve. The red line represents the best fit exponential to the All TCE bars. We select the threshold (8.68) where the average light curve will contain 0.2 TCEs.  \label{fig:TCE}}
\end{figure}

\begin{figure*}[!ht]
\centering \hfill \includegraphics[height=6.4cm]{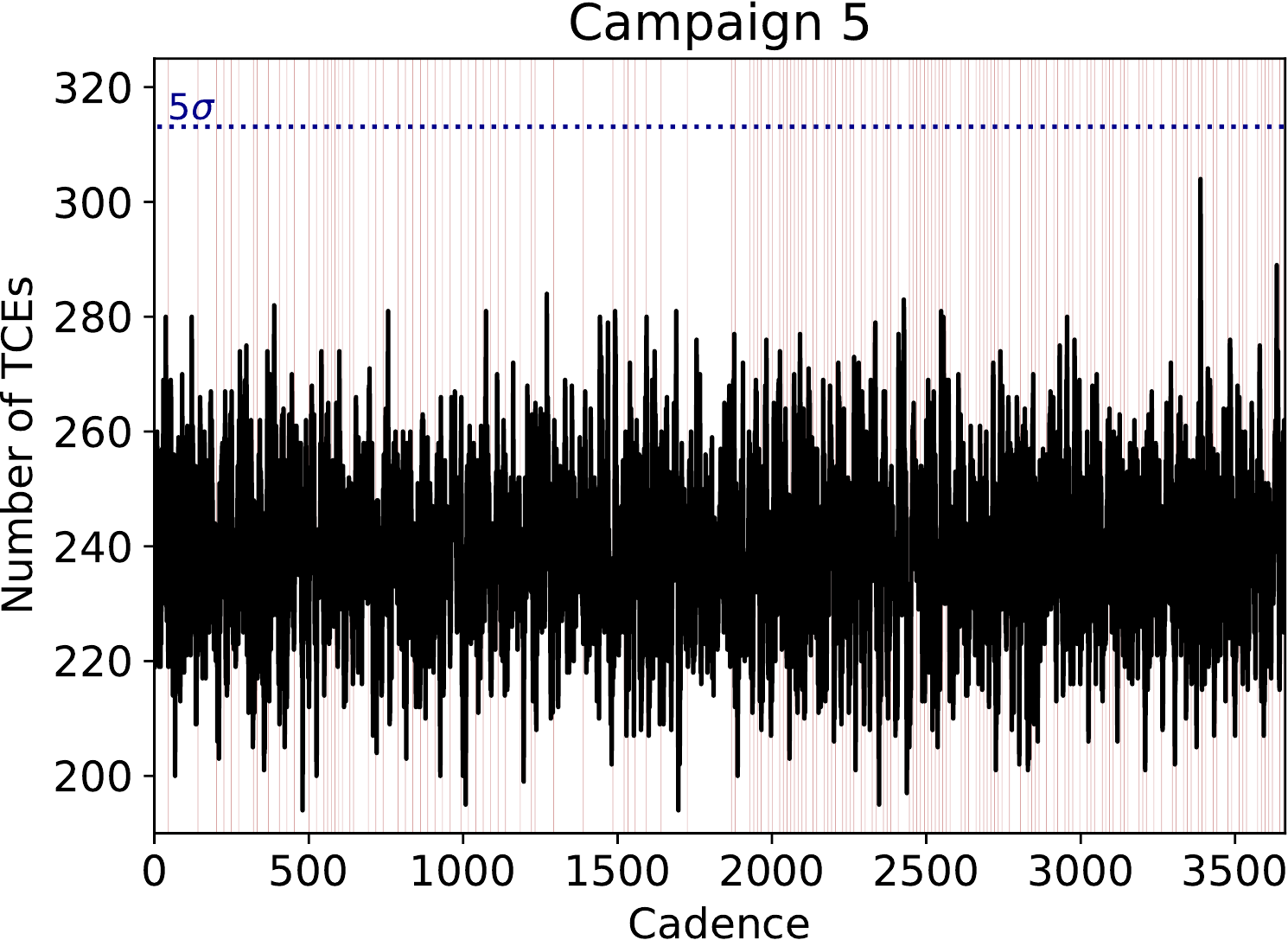}
\centering \hfill \includegraphics[height=6.4cm]{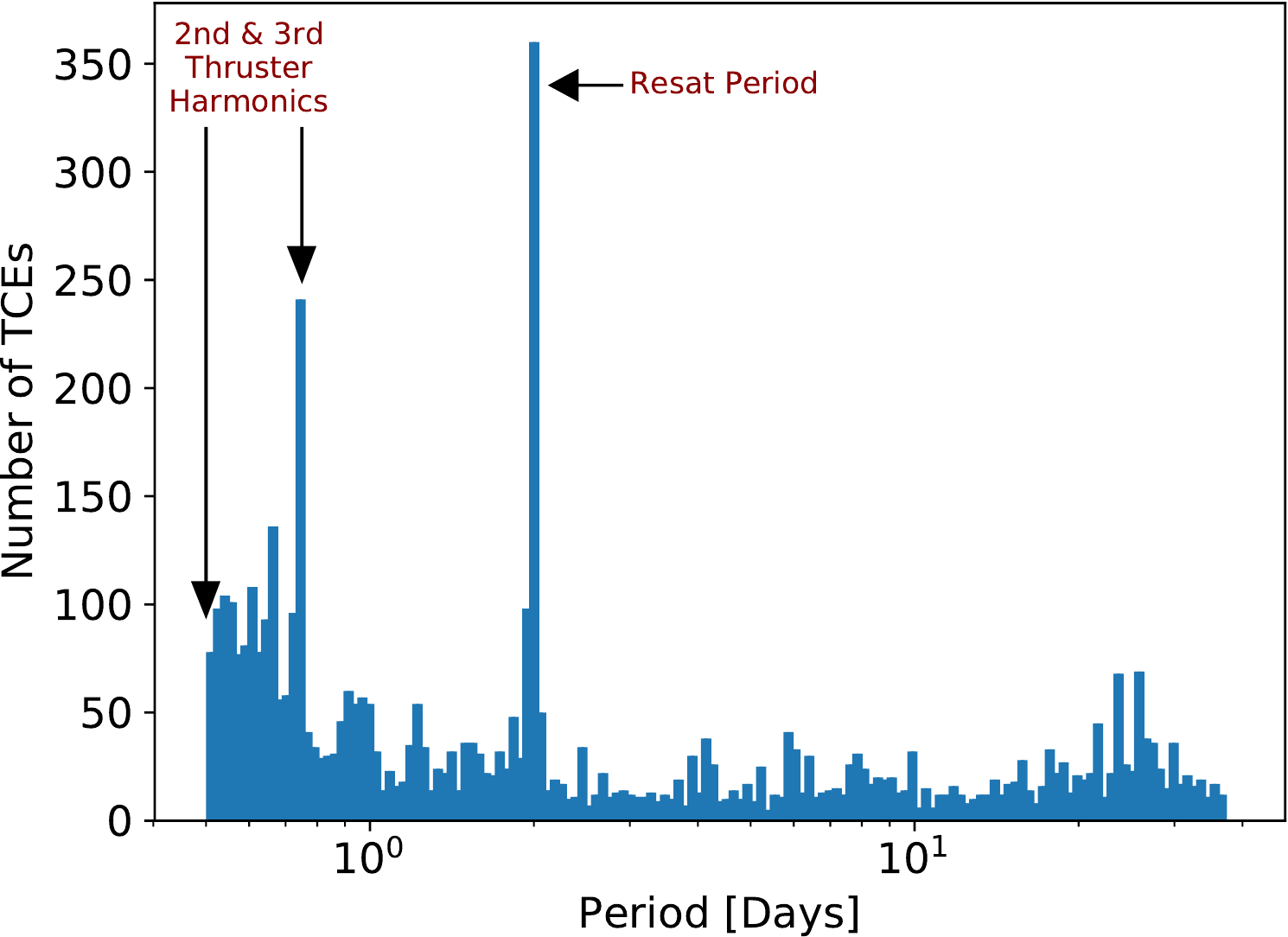}

\caption{\textbf{Left} The number of TCEs as a function of cadence. The blue dotted line represents $5\sigma$ above the median value. We find no cadences that exceed this limit, and therefore do not apply any additional masking. The red lines indicate the cadences in which a thruster firing occurred. \textbf{Right} shows the number of TCEs as a function of period. There are 150 uniform bins in log space. The outlier bins have been labeled to indicate their corresponding cause. \label{fig:sky}}
\end{figure*}

The \emph{Kepler} pipeline used a default TCE limit of $7.1\sigma$ \citep{jen02}. Any signal that produces a MES value of this magnitude or greater was considered for vetting. Assuming Gaussian statistics, this limit would only allow one false alarm signal through in the entirety of the \emph{Kepler} mission. However, the distribution of TCEs has long tails, where false alarm detections are more common at $7.1\sigma$ than originally expected. \emph{Kepler} found, on average, that 16\% of the targets produce a $\ge7.1\sigma$ event (198,706 light curves; 32,534 TCEs), and a majority of them were later classified as FPs \citep{tho18}.\footnote{In reality \emph{Kepler} detected TCEs in about 8\% of the light curves. A large fraction of these light curves contained multiple TCEs, inflating the average.} 

\emph{K2} photometry contains far more outliers than \emph{Kepler} prime photometry, so we tested different MES thresholds to look at the occurrence of TCEs at each limit (Figure \ref{fig:TCE}). The number of TCEs appears to match well with an exponential distribution, and the multiplicity of TCE occurrence does not appear to be affected by a difference in MES threshold. The number of ``First TCE'' detections, where we only consider the first TCE trigger for each light curve, is roughly 50\% of the ``All TCE'' detections at each of the tested thresholds. This indicates that a light curve where a TCE was triggered, was likely to trigger more than one TCE. If this ratio changed significantly at each tested threshold, an argument could be made to select a threshold at some level of similarity between the two values. Since no clear convergence occurs, we select a threshold (from our exponential fit) in which on average 20\% of the light curves will contain a TCE ($8.68\sigma$; determined by the fit exponential function). Since our threshold is higher than the \emph{Kepler} $7.1\sigma$, we sacrifice some sensitivity in order minimize the number of false alarm signals (\citet{tho18} found that reliability significantly increased for signals with MES$>8$ for the \emph{Kepler} pipeline). Previous studies have selected even higher thresholds \citep[$9\sigma$;][]{kru19}, which indicates that our survey will be able to detect smaller and noisier signals at the cost of a higher false positive rate. 

\subsubsection{Skye Excess TCE Identification}
\label{sec:skye}

One test implemented by \citet{tho18} was the ``Skye'' metric.
Inspired by this metric, we perform a similar analysis, in which we counted up the number of TCEs ($N_{\text{TCE}}$) found at each cadence (cadences considered detected are within the transit duration of the TCE). If all cadences contributed equally to the correlated noise in the light curves, the number of TCEs found at each cadence should be uniform. Any cadences that trigger an abnormally large $N_{\text{TCE}}$ were likely faulty and likely to cause FPs, warranting a global removal of this cadence from the time series. Such cadences have been found to be prolific in the \emph{K2} data set, as systematic outliers (due to the pointing jitters) commonly cause false TCE triggering. By manually tuning the threshold across different detrending software, found a $5\sigma$ threshold provided the best clipping. Thus, we considered a cadence faulty if $N_{\text{TCE}}-\text{med}(N_{\text{TCE}})>5\sigma_{N_{\text{TCE}}}$, where $\sigma_{N_{\text{TCE}}}$ is measured using the MAD estimator in Equation \ref{eq:MAD}. We find no cadences that exceed this limit in Campaign 5, but in a preliminary study we found several faulty cadences in the {\tt K2phot} Campaign 5 light curves \citep{pet15}. It appears that the {\tt EVEREST} software is proficient in removing such cadences. However, we will continue to search for faulty cadences in future campaigns. Figure \ref{fig:sky} shows the number of TCEs per cadence indicating a near-constant distribution. 

The number of TCEs produced at each period are shown in Figure \ref{fig:sky}. Two points clearly stand out: 0.75 days and 2 days. These two periods correspond to the frequency of various operations carried out by the spacecraft \citep{cle16}. 0.5 days and 0.75 days correspond to the 2nd and 3rd harmonics of the thruster firing, which occurs every six hours. Few TCEs are observed at the 2nd harmonic because it sits on the minimum period threshold. Some fraction of the potential TCEs would have been detected below 0.5 days and are now being detected at the next harmonic of 0.75 days. This operation re-aligned the telescope after drift caused by solar radiation pressure. Although {\tt EVEREST} masks these data points, the neighboring cadences can cause outliers. Thus, the harmonics of 0.25 days lead to an over abundance of TCEs where these outliers all line up. The jump at two days in Figure \ref{fig:sky} corresponds to the frequency of reaction wheel momentum resaturations (known as the ``Resat'' period). Although many of the cadences were masked by {\tt EVEREST}, the surrounding cadences can, again, cause a TCE. Our outlier vetting metric is proficient in removing these artificial signals (see Section \ref{sec:out})

\subsection{Prioritizing Reliability Over Completeness}
We selected parameters for our pipeline which maximize the completeness and the reliability of our sample. Adopting the philosophy of the \emph{Kepler} DR25 catalog \citep{tho18}, we willingly allow some known planet candidates to achieve a FP label in order create a uniformly vetted, highly reliable catalog. The parameters described in Section \ref{sec:vetter} have been tuned to minimize such misclassification, without allowing additional FPs through. We hope to maintain these same parameters across all campaigns, so that an aggregate occurrence calculation using all 18 campaigns can be made seamlessly.

\section{EDI-Vetter} \label{sec:vetter}

\begin{figure}
\centering \includegraphics[height=7.0cm]{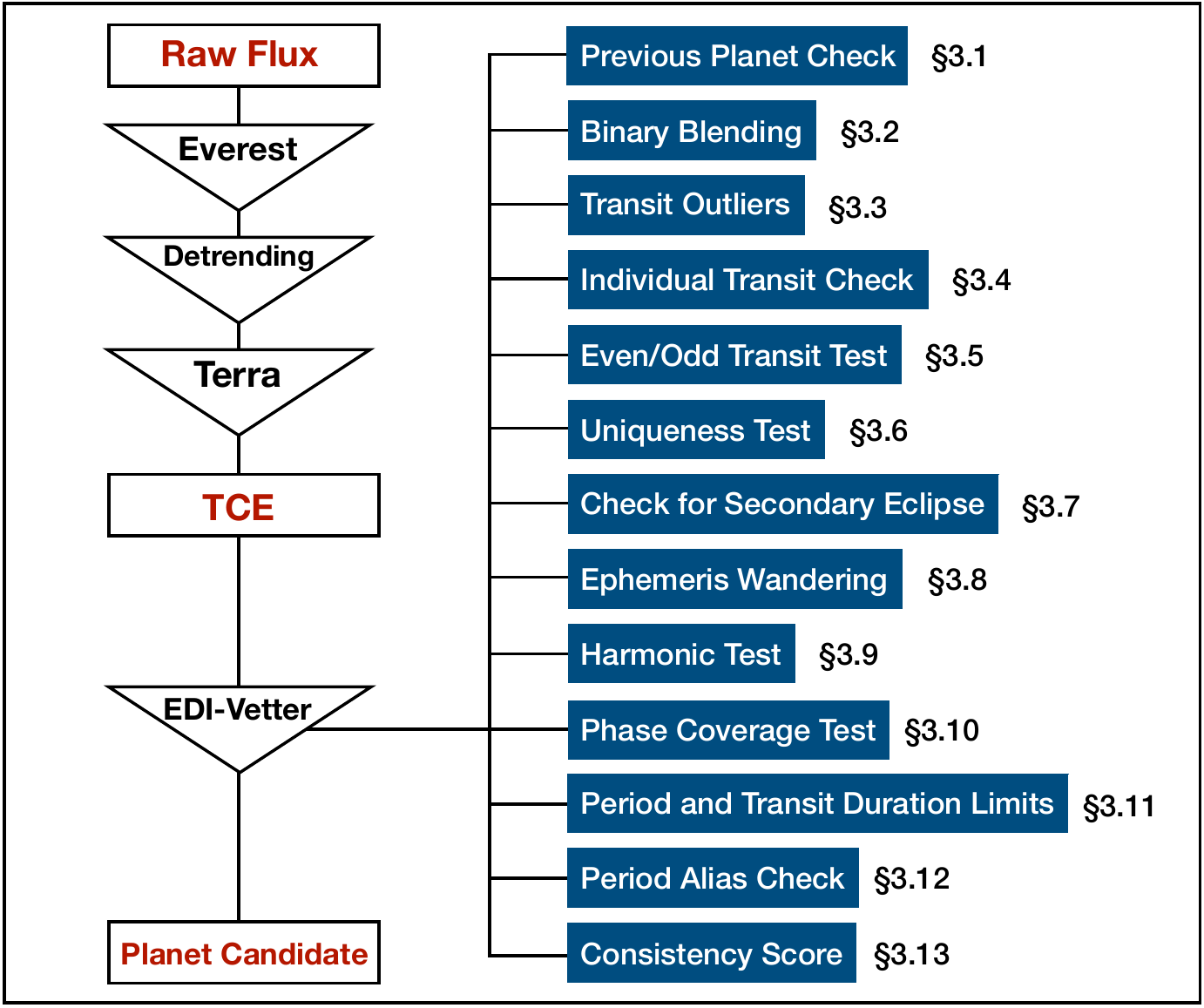}

\caption{A schematic of our \emph{K2} pipeline. The section numbers correspond to the detailed description of the {\tt EDI-Vetter} metric.  \label{fig:dia}}
\end{figure}

One of the more difficult tasks in any automated transit search is the vetting process. This is the final defense against transit FPs and must be robust against many different types of light curve anomalies. Automatted vetting processes have been created for the \emph{Kepler} pipeline \citep{mcc15,tho18}, but the \emph{K2} data set is far noisier and additional tests are required. We build upon the criteria used by the {\tt RoboVetter} \citep{tho18} to create {\tt EDI-Vetter} \citep{edi19}, which is more robust to the issues unique to the \emph{K2} data set. A user-friendly version of this software is made publicly available on GitHub\footnote{\href{https://github.com/jonzink/EDI-Vetter/}{https://github.com/jonzink/EDI-Vetter/}}. Here we discuss the metrics {\tt EDI-Vetter} uses to ensure our candidate list is pure with high reliability.

To begin, our software uses the maximum likelihood parameters found by {\tt TERRA} to fit a transit model to the detected TCE signal. Using the affine invariant sampler of \citet{goo10}, as implemented in Python by \citet{for13} ({\tt emcee}), and the {\tt batman} transit model \citep{kre15,man02}, we confirm that the {\tt TERRA} values found a global maximum likelihood and determine the maximum likelihood of additional transit parameters. We produce our transit model with the {\tt batman} Python package \citep{kre15}. Using 100 semi-independent walkers, 250 burn-in steps, and 100 parameter test steps (35,000 total step; 10,000 test samples), we estimate the maximum likelihood and uncertainty for the transit emphemeris ($t_0$), the radius ratio ($R_{pl}/R_{\star}$), the impact parameter ($b$), the signal period ($P$), the semi-major axis to stellar radius ratio ($a_{pl}$), the transit duration ($t_{dur}$), and the transit depth. We assume a circular orbit and derive two quadratic limb darkening coefficients using the stellar parameters available on \href{https://exofop.ipac.caltech.edu}{ExoFOP} \citep{hub16} along with the ATLAS model table for \emph{Kepler} bandpass limb darkening coefficients from \citet{cla12}. According to the literature for {\tt emcee} \footnote{\href{https://emcee.readthedocs.io/en/latest/tutorials/autocorr/}{https://emcee.readthedocs.io/en/latest/tutorials/autocorr/}}, 50 times the integrated auto-correlation time ($\tau$) is the lowest limit in which the MCMC sampler results can be trusted. We find $\tau\sim50$ samples for our 5 parameter model, indicating we should require at least 2,500 samples before our results can be considered meaningful. Clearly, the 10,000 samples taken by our MCMC is near this lower limit and likely insufficient for a thorough parameter estimation, but we sacrifice some accuracy to significantly improve the computational speed of our software. For vetted candidates we run a more thorough parameter estimation as described in Section \ref{sec:candidates}.

In the next 13 sub-sections (\ref{sec:previous}-\ref{sec:consistency}) we provide a detailed description of the filters used by {\tt EDI-Vetter} as shown in Figure \ref{fig:dia}. 

\subsection{Previous Planet Check}  \label{sec:previous} 
This test ensures that the detected TCE is not a repeat of a previous signal \citep[Section 3.2.2 of][]{cou16}. TCE re-occurrence is common for FPs, where masking is inefficient at removing the entirety of the light curve anomaly. Additionally, true signals which have periods misidentified by some integer multiple will produce multiple TCE signals. This test helps remove such redundancies. If the pipeline finds more than one TCE for a given light curve, this test will be enacted in decreasing signal detection order, i.e. the first signal detected will not be subject to this test, the second signal will be tested against the first, and the third signal will be tested against the first and the second signal.

We define $\Delta P$ as the separation in periods, normalized by the smaller of the two periods: 
\begin{equation}
\Delta P = \frac{P_B -P_A}{P_A}
\end{equation}
where $P_A$ and $P_B$ represent the shorter and longer periods respectively. To then determine the offset from the nearest integer multiple $\delta P$ is calculated. 
\begin{equation}
\delta P = |\Delta P - \textrm{round}(\Delta P)|
\end{equation}
where the round() function will round to the nearest integer value. We can then determine how statistically significant ($\sigma_{P}$) a period separation is using the erfcinv() (inverse complementary error function)
\begin{equation}
\sigma_{P}=\sqrt{2}\textrm{ erfcinv}(\delta P)
\end{equation}
where larger $\delta P$ values will produce small $\sigma_{P}$ values. For consideration as a candidate we require high detection order signals (signals found after the first signal within the light curve) to be separated with a $\sigma_{P}\le2$ for candidate consideration. If a signal matches with a previous detection with a $\sigma_{P}>2$, it is likely a repeated detection. However, we also must consider the case of resonant systems, where integer period separations are common. To avoid falsely eliminating such systems, we also consider the ephemeris separation for the two signals ($\Delta t_0$).
\begin{equation}
\Delta t_0 = \frac{|t_{0A} -t_{0B}|}{t_{dur}}
\end{equation}
where $t_{dur}$ is the transit duration of the signal in question and $t_{0A}$ and $t_{0B}$ are the ephemeris times for the first transit within the data set. If $\Delta t_0$ is large ($\ge1$) and $\Delta P$ is small ($\le2$), it is possible that the signals are from a resonant orbit. The caveat to this is the case where the higher detection order signal is fitting to the secondary eclipse of the transit. To ensure this is not the case $\Delta SE_1$ and $\Delta SE_2$ are calculated:

\begin{equation}
\begin{aligned}
    \Delta SE_1 = \frac{|t_{0A} -t_{0B}+P_A/2|}{t_{dur}} \\
    \Delta SE_2 = \frac{|t_{0A} -t_{0B}-P_A/2|}{t_{dur}}
\end{aligned}
\end{equation}
Again, a small $\Delta SE_1$ or $\Delta SE_2$ ($<1$) indicates the higher order detection is likely fitting to the secondary eclipse. This metric will only be sensitive to circular orbit eclipsing binaries, where the secondary eclipse is at a phase of 0.5. Future iterations of this pipeline will consider non-circular orbits in this metric.

All of these metrics are assuming the previous planet was a real signal. If the previous detection was deemed a FP and $\sigma_{P}>2$, this signal is likely just a repeated detection from an under masked anomaly. Thus, the signal will be flagged as a FP.

For clarity, we summarize here the full process in which a signal will be classified as a FP. If $\sigma_{P}>2$ and the previous signal was deemed a FP, this detection will be flagged as a FP. If $\sigma_{P}>2$ and $\Delta t_0<1$, this detection will be flagged as a FP. If $\sigma_{P}>2$ and $\Delta SE_1<1$ or $\Delta SE_2<1$, this detection will be flagged as a FP. Otherwise, this signal will continue to be considered for candidacy. 

\subsection{Binary Blending}
\label{sec:blend}
In this metric we seek to identify cases where the signal is due to an eclipsing binary, either from the target or from a nearby source, but with such a large impact parameter and/or dilution from the third light source that the observed depth is comparable to that of a transiting planet. During the original \emph{Kepler} mission, flux contamination from nearby sources was ruled out by fitting the pixel response function model \citep{bry10} to the star in and out of transit, looking for offsets in the photo-center of the light (the centroid). If statistically significant differences appeared, the target was considered contaminated \citep{mul17}. Two scenarios could result in such a shift. First, contamination from a source within a few arcseconds of the primary target would produce this type of shift during transit. Second, a very deep transit signal from a nearby eclipsing binary could contaminate a static light curve, causing a shift during transit. Such methods are difficult to apply to \emph{K2} because of the roll experienced by the telescope, which moves the target image across several pixels. Using cadences with similar roll angles, the {\tt DAVE} software is able to recoup some of this detection ability \citep{kos19}. However, the shorter data span of each \emph{K2} campaign, and the lack of similar roll angle data, makes finding these statistical differences more difficult. Instead, we use the Gaia DR2 data set \citep{gai18} to look for flux contamination from nearby sources. 

Gaia can resolve nearby sources down to $1\arcsec$ for G-band $\Delta \mathrm{mag}\lesssim3$ \citep{zie18}. Since the \emph{K2} pixel scale is $3.98\arcsec$, we can achieve sub-pixel resolution using cross-matching alone. We utilize the gaia-kepler.fun\footnote{\href{https://gaia-kepler.fun}{https://gaia-kepler.fun}} cross-match database with a $20\arcsec$ search radius to look for these contaminants. To ensure we have correctly matched the target star, we require the \emph{``phot\_g\_mean\_mag''} parameter to be within 1 magnitude of the assigned \emph{``k2\_kepmag''}. These values are often almost indistinguishable because the \emph{Gaia} G band and \emph{Kepler} band probe nearly the same wavelength range. If this criteria is not met, we check the next closest source within the $20\arcsec$ search radius to see if it meets this criteria. This process will continue until a target of the correct magnitude is found, or all of the sources in the $20\arcsec$ search radius have been tested. If no source is within this 1 magnitude limit, we will select the source that provides the closest magnitude match. This closest match scenario was applied to 301 of the C5 targets (1.2\% of the total tested targets). 

As used for the {\tt EVEREST} pipeline \citep{lug18}, we adopted aperture \#15 from the {\tt K2SFF} catalog \citep{van14}. Using the size of the aperture, we can determine the amount of flux contamination from neighboring binaries. Often the apertures select pixels in a circular manner around the target, but in cases where additional point sources are nearby, the aperture may be elongated, capturing additional flux from the nearby source. To account for this possible elongation, we consider flux contamination within this radius ($\Delta ap$):
\begin{equation}
\Delta ap = 3.98\arcsec\sqrt{\frac{N_{pix}}{\pi}+1}
\end{equation}
where $N_{pix}$ is the number of pixels for the given aperture. This $\Delta ap$ is likely overestimating the flux contamination for many of the well-behaved apertures, where the aperture is large and circular, but is important to ensure high reliability. To determine the amount of flux contamination, we compute the flux ratio for the nearest potential contaminating source ($F^{Ratio}_{Gaia}$):
\begin{equation}
\frac{F_{neighbor}}{F_{\star}}=F^{Ratio}_{Gaia}=10^{\frac{\Delta m_{Gaia}}{-2.5}}
\label{eq:rat}
\end{equation}
where $F_{neighbor}$ is the expected flux from the nearby source, $F_{\star}$ is the expected flux from the target star, and $\Delta m_{Gaia}$ is the difference in magnitude between the target source and the possible contaminant in the Gaia G band. To determine the impact of this external source on our target, we calculate the ratio of total flux to the target flux ($F^{Tot}_{Gaia}$):
\begin{equation}
\resizebox{.9 \columnwidth}{!}{%
$F^{Tot}_{Gaia}=\frac{F_{\star}+F_{cont}}{F_{\star}}=1+\frac{F^{Ratio}_{Gaia}}{2}\Bigg(1+\erf\bigg(\frac{\Delta ap - \Delta d}{2.55\arcsec\sqrt{2}}\bigg)\Bigg)$%
}
\label{eq:tot}
\end{equation}
where $F_{cont}$ is the fraction of flux from $F_{neighbor}$ within the aperture of the target and $\Delta d$ represents the angular distance between the target and the contaminating source.  Using the error function in this method assumes a 1D Gaussian PSF with a standard deviation of $2.55\arcsec$ (corresponding to the $6\arcsec$ FWHM of the \emph{Kepler} PSF). The 1D assumption will certainly overestimate the amount of flux contamination, but this cautious approach will improve our reliability against astrophysical FPs. For targets with multiple neighboring stars we consider the flux contamination from all the sources within a $20\arcsec$ radius of the source.

\begin{figure*}[!ht]
\centering \hfill \includegraphics[height=6.4cm]{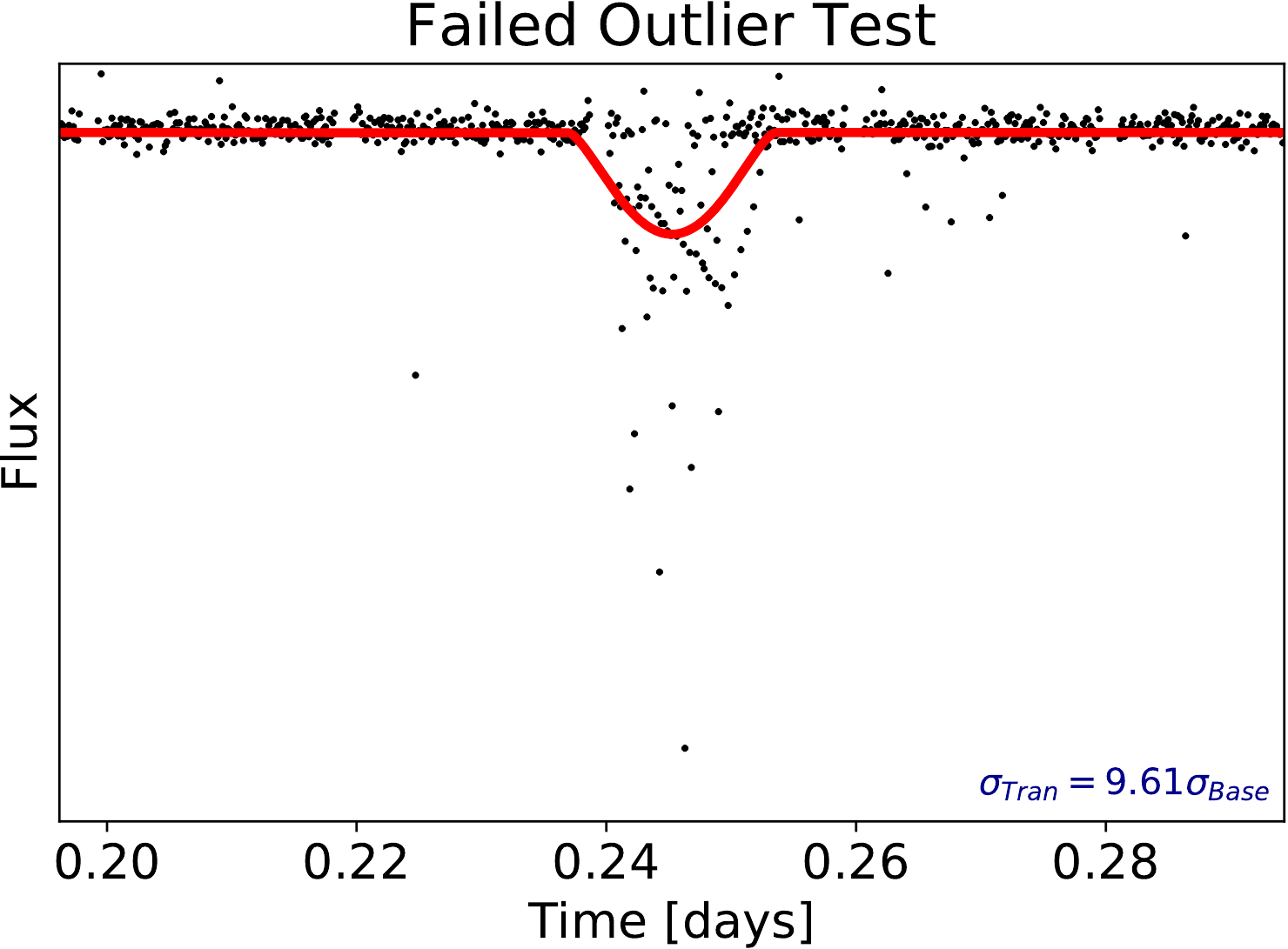}
\centering \hfill \includegraphics[height=6.4cm]{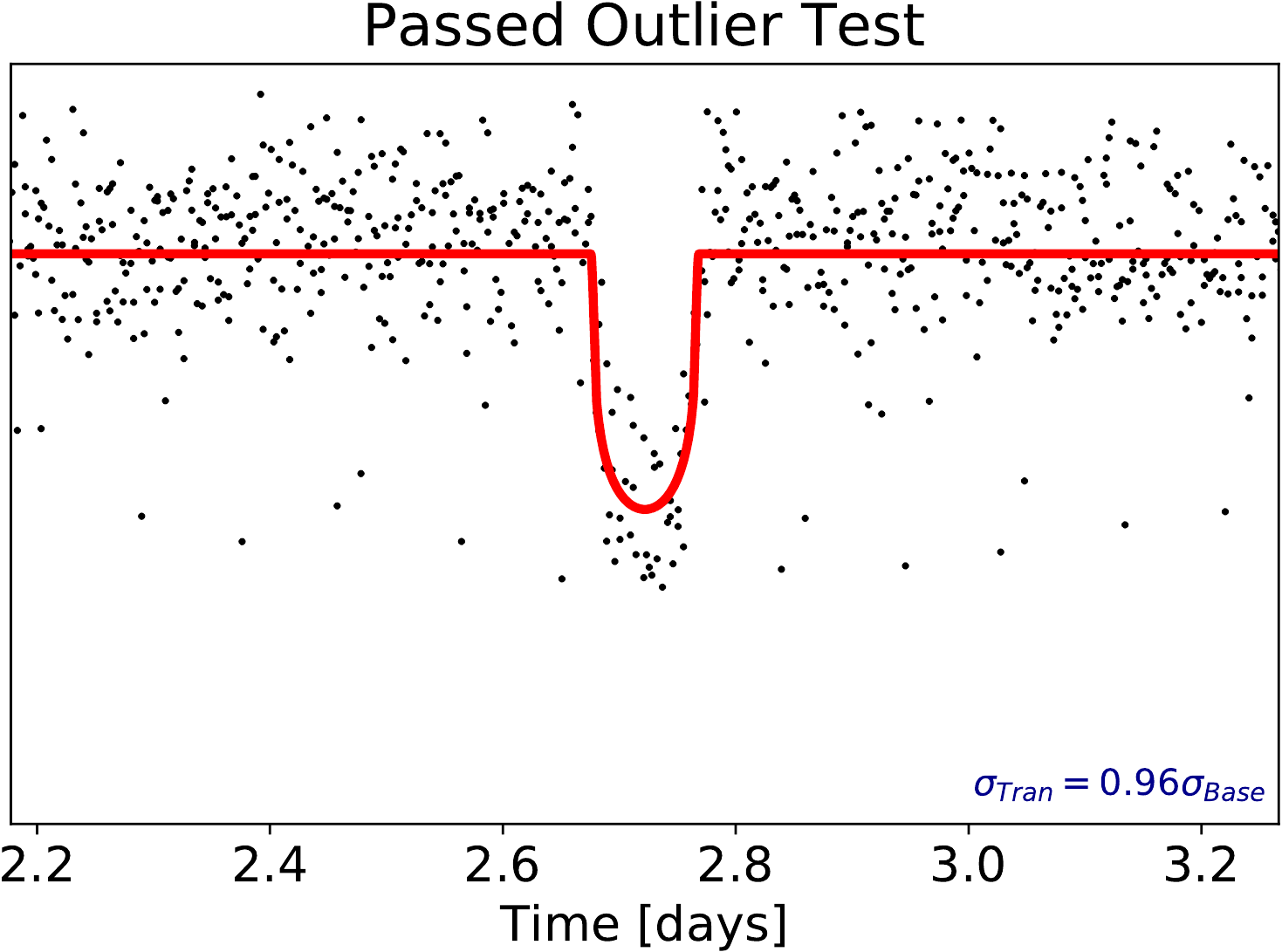}

\caption{\textbf{Left} A TCE with a period of 0.5 days, likely caused by the alignment of measurements made near thruster firings. The large in-transit RMS ($\sigma_{Tran}$) is an indicator that the singal is a FP. This TCE was rejected using the outlier variance test. \textbf{Right} EPIC 211331236.01 with a period of 5.4 days. The in-transit deviations match well with the out of transit fluctuations. Thus, this signal was not removed by our outlier variance test.  \label{fig:outliers}}
\end{figure*}

One limitation to this approach is that our search limit is bound by the $20\arcsec$ search radius of the \href{gaia-kepler.fun}{gaia-kepler.fun} cross-match. In some rare instances, an aperture will extend beyond this radius ($N_{pix} >78$). Upon visual inspection, most of these targets are in very crowded fields with several stars within a single aperture. Implementing this aperture limit, we remove 54 targets from our search. Additionally, contaminating sources can also exist beyond the $20\arcsec$ search radius. To help account for these cases, we also use the 2MASS J-band photometry. Pulling from ExoFOP\footnote{\href{https://exofop.ipac.caltech.edu}{https://exofop.ipac.caltech.edu}}, which provides 2MASS cross-matches beyond $20\arcsec$, we calculate equation \ref{eq:rat} for 2MASS and find the potential contamination ($F^{Tot}_{2MASS}$):
\begin{equation}
F^{Tot}_{2MASS}=1+\frac{F^{Ratio}_{2MASS}}{4}\Bigg(1+\erf\bigg(\frac{\Delta ap - \Delta d}{2.55\arcsec\sqrt{2}}\bigg)\Bigg)
\end{equation}
Here, it is important to note that we assume the flux contamination will be half as strong as a similar contaminant in the Gaia band. This is to account for the fact the 2MASS J-band photometry cannot be directly compared to the expected \emph{K2} flux. \citet{how12} suggests a flux ratio of $\sim1/4$ for a J-band to K-band conversion, however we choice $1/2$ to err on the side of over-estimating potential contamination. We find that $F^{Tot}_{2MASS}$ is usually very close to unity, only in extreme cases where the J-band contamination is significant, and $F^{Tot}_{Gaia}=1$, will the 2MASS flux ratio play a role.

To determine the impact of this flux contamination on our source, we consider how $F^{Tot}_{Gaia}$ and $F^{Tot}_{2MASS}$ affect the inferred planet radius. One of the biggest concern with binary contamination is the potential to decrease the transit depth and falsely present an eclipsing binary as a planet candidate. Additionally, in systems with a transiting planet, the planet radius is derived from the depth of the measured transit. When flux from an additional star is present, the depth of the transit will be diluted and result in an underestimation of the planet radius \citep{cia17,ful17,mat18}. We adopt the same radius and impact parameter requirements suggested by \citet{bat13}, but now include a flux contamination correction.
\begin{equation}
\frac{R_{Pl}}{R_\star}\sqrt{\textrm{max}(F^{Tot}_{Gaia},F^{Tot}_{2MASS})} + b\le 1.04
\label{eq:EB}
\end{equation}
where $b$ is the impact parameter and $R_{Pl}/R_\star$ is the radius ratio fit to the transit in question. The max function chooses the higher of the two flux ratios. This ensures we do not double count the flux contamination from a source. The median $F^{Tot}$ value for Campaign 5 targets is 1.00003, indicating that most targets are well isolated. Any signals that exceed the provided limit (Equation \ref{eq:EB}) will be flagged as a FP. This method of FP flagging allows us to remain agnostic to the stellar parameters. Additionally, we put an upper limit on $R_{Pl}/R_\star$. Any TCE found with $R_{Pl}/R_\star>0.3$ will automatically be flagged as a FP. This limit is imposed to eliminate low impact parameter ($b$) eclipsing binaries.

\subsection{Transit Outliers}
\label{sec:out}

One common issue seen in many of the TCEs found in the \emph{K2} light curves is the chance alignment of several outliers. This metric is aimed to identify when such cases occur. The general increase in noise caused by the spacecraft roll motion and thruster firing caused a significant number of outliers to occur in the flux data set. Much of this is removed by the {\tt EVEREST} processing, but occasionally a few will make it through (evidence of this can be seen in the right-hand panel of Figure \ref{fig:sky}). When this occurs, the TCE search is likely to find a period where the outliers line up when phase folded. These chance alignment signals are hard to distinguish with previously developed vetting tools, but easy to spot by eye. Here we present a unique test to eliminate these anomalies.

We use the best fit model to subtract the signal from the folded data set. We first check to see if the in-transit residuals have a larger variance ($\sigma_{\textrm{Tran}}$) than the rest of light curve ($\sigma_{\textrm{Base}}$). However, this calculation is complicated by larger MES signals. We found that many high MES transit signals will produce larger variances in the residuals because of either poor fitting or poor detrending. Fortunately, many of these chance alignment outliers produce a low MES signal. Thus, we deem any transit residual with $\sigma_{\textrm{Tran}}>(0.461*\textrm{MES}-2)\sigma_{\textrm{Base}}$ a FP. This threshold was tuned in correspondence to an acceptable increase of $2\times$ the in-transit RMS at the detection limit and $4\times$ the RMS at MES$=13$. In Figure \ref{fig:outliers} we show two signals, one which passed this test and one which failed.   

We often find that several aligned outliers will trigger a TCE, but retain a low $\sigma_{\textrm{Tran}}$ due to a larger number of non-outlier baseline flux measurements included in the $\sigma_{\textrm{Tran}}$ calculation. To account for these cases, we count up the number of in-transit cadences that produce outliers (residual$>3\sigma_{\textrm{Base}}$) in the transit residuals. If this number ($Out$) exceeds the following limits, the signal is deemed a FP:
\begin{equation}
\begin{aligned}[b]
\textrm{if MES}<40:\;\;\;  &  Out>N_{\textrm{Tran}} \\
\textrm{else}:\;\;\;  &  Out>0.3\times\textrm{MES}+4
\end{aligned}
\label{eq:tran}
\end{equation}
where $N_{\textrm{Tran}}$ is the number of transits. Equation \ref{eq:tran} allows, on average, one outlier per transit for signals with MES$<40$. For stronger signals, outliers are more likely to be triggered due to poor model fitting. This problem becomes more significant as MES increases, thus we use the linear function of MES as a threshold for these cases. This function was tuned to allow 16 outliers at MES$=40$ and and 25 outliers at MES$=70$. This metric performs well against outlier alignments, but has a mild tendency of flagging high MES, but poorly detrended planets, as FP. This is the unfortunate cost of attaining a high reliability catalog.

\subsection{Individual Transit Check}
\label{sec:tran}
Much of this part of our pipeline is derived from the ``Marshall'' test \citep{mul16}. For clarity we will briefly describe how this test works. We look at each transit within the light curve individually. Each the transits detected are then fit to four potential models:
\begin{equation}
\resizebox{.9 \columnwidth}{!}{$
\begin{aligned}[b]
f(t) & =y_0 & \textrm{Flat}\\
f(t) & =y_0 + S(t) & \textrm{Logistic}\\
f(t) & =y_0 + S(t)(1 - e^{\beta t}) & \textrm{Logistic-exponential}\\
f(t) & =y_0 +S(t-\tau/2)-S(t+\tau/2) & \textrm{Double-logistic}
\end{aligned}
$}
\end{equation}
where $y_0$ is a constant offset and $\tau$ and $\beta$ are tunable parameters (see Figure 1 of \citet{mul16} for reference). We also do not fit an additional GP (as noted in the original Marshall test) because the light curves have already experienced several GP fitters. The function $S(t)$ is a logistic function given as:
\begin{equation}
S(t) = \frac{d}{1+e^{\gamma(t-t_0)}}
\end{equation}
where $t_0$, $\gamma$, and $d$ are all tunable parameters. These functions are meant to replicate possible FP signals. To gauge the success of each fit, we calculate the Bayesian Information Criterion (BIC) for each function. This value is meant to mimic the Bayesian evidence of a Gaussian likelihood, and penalizes for increasing the number of tunable parameters. We also calculate the BIC for our best fit transit model. If the BIC value for any of the previously noted functions is 10 less than the transit BIC ($BIC_{f(t)}+10<BIC_{\textrm{Transit}}$) and the MES/$\sqrt{N_\textrm{Tran}}>4$ (this criterion ensures the expected single transit signal strength is large enough to provide a meaningful model fit), we mask that transit signal. However, the MES/$\sqrt{N_\textrm{Tran}}>4$ requirement is ignored for the constant $y_0$ model.

We also consider the number of unmasked points in each transit. If more than 50\% of cadences within the transit duration are masked, the entire transit is masked. This ensures that a significant fraction of the transit data is available. Here we treat all parts of the transit signal equally, but in reality the egress and ingress are more important for candidacy. We consider this further in Section \ref{sec:phase}. The current test has the potential to mask several transits. To see how this affects the signal we reanalyze the light curve to see if the transit still has three observable transits and a MES $\ge 8.68$. If the signal now fails to meet either of these requirements, the signal is flagged as a FP. 

Looking at the Single Event Statistics (SES) of each transit, we can determine if a single transit is dominating the MES. In an ideal case the SES=MES/$\sqrt{N_\textrm{Tran}}$. {\tt EDI-Vetter} will look for SES values that exceed 80\% the overall MES value. In the scenario where only 3 transits exist we would not expect any transit to exceed $\sim60\%$, thus 80\% provides an unlikely natural occurrence. As performed in Section 3.2.4 of \citet{cou16} and Section A.3.5 of \citet{tho18}, this test will help eliminate single transit outliers that pushed the MES over the TCE limit. It is not robust to chance alignments of multiple transit outliers as addressed in Section \ref{sec:out}. If this limit (SES$>0.8$MES) is exceeded, the signal is deemed a FP by the pipeline.

One potential issue with this test is that a true planet transit which falls on a systematic feature could inflate the SES beyond this threshold, triggering a FP flag. In testing this was not an apparent issue, but something we will continue to monitor as we move forward with the other campaigns. 

\subsection{Even/Odd Transit Test}
\label{sec:evenodd}
This classic test will look for cases where the transit primary and secondary eclipse are folded on top of each other. This will occur when the secondary eclipse depth is a significant fraction of the primary eclipse depth and the orbit is circular so that the secondary eclipse occurs at a phase of 0.5. We would only expect such similarity between transit depths for eclipsing binaries. Thus, when we find over-folded light curves of this nature, we flag them as a FP.

To perform this test, we separate the phase-foldings into even and odd transits (i.e. every other transit in one group and the remainder in the other group). The transits are then refit, holding $P$ constant. We compare the inferred $R$ (radius ratio) values from each group (odd vs. even) to look for statistical differences ($Z$).
\begin{equation}
Z = \frac{R_{even}-R_{odd}}{\sqrt{\sigma_{R_{even}}^2-\sigma_{R_{odd}}^2}}
\end{equation}
where the $\sigma$ values correspond to the labeled $R$ fit uncertainties. We select $Z>5$ for FP flagging.

One notable issue arises when the true period of the signal is twice the period detected, folding baseline data on top of the transit signal. This will be flagged as FP despite being a true planet signal. Cases where this occurs will be identified in Section \ref{sec:alias} and re-run at the correct period folding.

One of the main complications of this test are the cases where one of the transits (usually near the end or beginning of the campaign) are poorly detrended. If this occurs in a true signal with very few transits ($N_{Tran}\le5$), one of the two groups can be significantly pulled by this faulty transit and cause this flag to be falsely triggered. Many of the known high-MES candidates not found by our pipeline fall victim to this flagging. This problem can be easily spotted by eye, but remains difficult to automate.

\subsection{Uniqueness Test}
\label{sec:unq}

This metric identifies cases where the phase folded light curve appears to produce several transit-like dips. This type of noisy light curve is unlikely to reveal a real planet signal, but rather highlight the systematic issues of the spacecraft. We base this metric on the ``model-shift uniqueness test'' \citep{row15, mul15, cou16, tho18}. Here, we provide a brief explanation of this test. For a thorough explanation, we refer interested readers to \citet{cou17b}.

The first step in this process is to fold the light curve according to the period of the signal in question. The light curve should remain folded on this period for the entirety of this test. To determine how unique a signal is to the light curve, it is first important to mask the detected signal from the light curve and any possible secondary eclipse. This done by masking the largest signal (MES$_1$) and the second largest signal (MES$_2$) in the light curve, given that the second largest signal is within 50\% of the transit duration ($t_{dur}$) of the expected secondary eclipse location. If the signal MES$_2$ is not within the expected secondary eclipse window, the third largest signal (with the same $t_{dur}$) is assigned to MES$_2$ before masking. This allows the second largest signal to remain in the light curve for detection as MES$_3$.  We can then measure the red noise within the period folded data. This is achieved by fitting a transit model to each flux point with a fixed width of the original signal duration ($t_{dur}$). We simplify this process by using a box shape transit, where the best-fit depth can be analytically shown as the mean of the selected points. This simplification significantly increases the speed of our processing. If the light curve has no red noise, these averaged depth values will have a standard deviation ($\sigma_{Red}$) on the order of the light curve noise ($\sigma_{LC}$), otherwise it will be much larger. We then calculate the ratio of these two values ($F_{Red}$):
\begin{equation}
F_{Red}=\sigma_{Red}/\sigma_{LC}
\end{equation}

Additionally, we want to determine the strength of the third largest signal (MES$_3$) within the light curve, and then the largest flux brightening signal (the largest signal found when the light curve is inverted; MES$_4$).

For statistical comparison, we consider the threshold in which a signal is statistically significant under the assumption of Gaussian noise:
\begin{equation}
F_1=\sqrt{2}\textrm{ erfcinv}\Bigg(\frac{t_{dur}}{P\times N_{TCE}}\Bigg)
\end{equation}
where $P$ is the period in which the light curve has been folded, and $N_{TCE}$ is the number of TCEs expected in the entire campaign. In an effort to ensure high reliability, we select $N_{TCE}=25,000$, or roughly one TCE per light curve. Here $P/t_{dur}$ represents the number of events in each light curve and $N_{TCE}$ is the number of light curves considered. For the statistical significance within a single light curve, we consider the threshold as follows:
\begin{equation}
F_2=\sqrt{2}\textrm{ erfcinv}\Bigg(\frac{t_{dur}}{P}\Bigg)
\end{equation}
We consider a signal viable for further vetting if all the following inequalities are satisfied:
\begin{equation}
\begin{aligned}[b]
 F_1 - (\textrm{MES}_1/F_{red}) & \le 1\\
 F_2 - (\textrm{MES}_1-\textrm{MES}_3) & \le 2\\
 F_2 - (\textrm{MES}_1-\textrm{MES}_4) & \le 3\\
\end{aligned}
\end{equation}
If any of the previous equations are false, the signal is flagged as a FP (see Figure 19 of \citet{cou17b} for reference). The inequality threshold values are based on those provided in \citet{tho18}, but have been manually tuned to suit the needs of this pipeline.  

We also consider the significance of the secondary eclipse detected within the folded light curve. If the phase of the secondary eclipse falls within $0.5t_{dur}$ of $t_0+P/2$, and $\textrm{MES}_1-\textrm{MES}_2-F_2>0$, the secondary eclipse appears genuine and warrants further testing in Section \ref{sec:SE}. This signal will be flagged with a secondary eclipse flag. It is important to note that this test will only be sensitive to companions with near circular orbits, where the secondary eclipse occurs at or near a phase of 0.5. Future iterations of the pipeline will consider eccentric orbits. If the secondary eclipse does not fall at or near a phase of 0.5, the signal is deemed a FP due to lack of uniqueness if all of the following inequalities are satisfied:

\begin{equation}
\begin{aligned}[b]
\label{eq:uniSE}
 \textrm{MES}_2/F_{red}-F_1 & > 1\\
 \textrm{MES}_2-\textrm{MES}_3-F_2 & > 0\\
 \textrm{MES}_2-\textrm{MES}_4-F_2 & > 0\\
\end{aligned}
\end{equation}
If any of the previous equations are false, the signal will be further vetted for candidacy. If the secondary eclipse is not astrophysical, the inequalities in Equation \ref{eq:uniSE} make certain that the original MES$_1$ signal is still sufficiently unique.

\subsection{Check for Secondary Eclipse}
\label{sec:SE}
Apparent secondary eclipses are often the hallmark of an eclipsing binary. However, such events can also be seen for certain transiting hot Jupiter-like planets. In the previous section, we discussed the criteria for a secondary eclipse to be considered statistically significant. If the TCE was flagged as a secondary eclipse candidate, we consider several metrics before determining how to classify the signal.

Inspired by the frame work of Section A.3.2 in \citet{tho18}, we begin by fitting a transit model to the secondary eclipse in question. If the secondary eclipse model depth is greater than 10\% of the depth of the initial transit, the secondary eclipse radius ratio is statistically significant ($R_{SE}>2\sigma_{R_{SE}}$), and the initial transit impact parameter $b\ge0.8$, the TCE is flagged as a FP. The large secondary eclipse and high impact parameter are indicators that the transit is likely a grazing eclipsing binary. If the signal fails to meet any of the listed criteria, the TCE remains labeled as a secondary eclipse candidate, and continues for further vetting. Such cases are likely transiting hot Jupiters reflecting a significant amount of light. It is possible that a well aligned, low impact parameter eclipsing binary could potentially slip through this metric, but the occurrence of such an event would likely produce a large enough transit depth to trigger the metrics discussed in Section \ref{sec:blend}.     

\subsection{Ephemeris Wandering}
The emphemeris value ($t_0$) found by the grid TCE search ({\tt TERRA}) will be very similar to the value determined by the transit model fitting if the signal is a true planet candidate. In cases where the two values differ significantly, the signal is likely very asymmetric and does not fit well with the transit model. To avoid these contaminants, we flag all signals as FP if the $|\Delta t_0|$ between the grid search and the model fitting is greater than $0.5t_{dur}$.         

\subsection{Harmonic Test}
\label{sec:Haromic}
Sinusoidal stellar variability is a common trigger for TCE search algorithms. In Section \ref{sec:har} we attempted to detrend such periodicity, but our limits of $P<0.5$ days for the harmonic fitter and $P>5$ days for the GP fitter create a window in which a sinusoidal signal can sneak in. Furthermore, cases where the sinusoidal signal is not completely removed can cause TCE triggering. To eliminate such contamination, we fit several sine curves to test for such FPs.

When testing the possible harmonic functions, we fit a sine curve to the data while holding fixed the period and allowing the phase and amplitude ($A_{sin}$) to vary. We attempt to fit six different harmonic periods: $P$ (the TCE period), $P/2$, $2P$, $t_{dur}$ (the transit duration), $2t_{dur}$, and $4t_{dur}$. These potential periods represent the most notorious harmonics found mimicking a TCE.

When considering how well the sine curves fit the data, it is important to only consider the $A_{sin}$ value. Any attempt to use the BIC value, as done in Section \ref{sec:tran}, can cause one to misclassify hot Jupiters, which can naturally produce strong sinusoidal oscillations within the folded light curve. If any of the tested periods produce $A_{sin}/\sigma_{A_{sin}}>50$, the signal is labeled a FP. In cases where the harmonic signal is caused by a hot Jupiter, the primary and secondary eclipse will reduce the sine signal strength below $50\sigma$. Furthermore, a large amplitude in it self may be an indicator that the oscillations are of stellar origin. Our second test checks to see if any of the $A_{sin}$ values are larger than 90\% of the fit transit depth and $A_{sin}/\sigma_{A_{sin}}>2$, ensuring the signal is real. If this criteria is met, the signal is classified as a FP. Large $A_{sin}$ values correspond to a harmonic that is contributing to a large fraction of the signal depth and is likely the cause of the TCE.

As mentioned in Section \ref{sec:har}, part of our pre-processing attempts to fit out harmonic trends in the light curve. To account for the possibility of artificially introducing a signal which could be detected as a TCE, we calculate the power (Pow$_{TCE}$) of a Lomb-Scargle periodogram at the transit period ($P$), and then search for the largest signal (Pow$_{max}$) with a period less than the transit period but greater than twice the cadence spacing. The goal here is to determine if the TCE is some integer multiplier of the initial harmonic fitter. If Pow$_{max}>($Pow$_{TCE}+0.5)$ and the Pow$_{max}$ signal is found with a period less than 0.5 days, the TCE is deemed a FP.  

\subsection{Phase Coverage Test}
\label{sec:phase}
With the significant processing required to detrend the \emph{K2} light curves, on average $\sim2\%$ of the data is masked out on any given light curve due to systematic issues causing flux outliers. Although the test performed in Section \ref{sec:tran} will remove transits that have less than 50\% phase coverage, it is still possible for phase folded data to contain large gaps in the transit signal. This occurs when all the masked portions of the light curve line up in phase space. This is especially common for planets with periods that are near integer multipliers of the \emph{K2} cadence. It is difficult to properly vet such cases, as the transit appears incomplete and prone to FP contamination. Thus, we have developed a metric for flagging TCEs with such gaps.

Chance outlier alignments lack an important trait, they do not have a pronounced ingress and egress. To capitalize on this feature, we require more phase coverage during these periods than during the transit mid-point. The phase folded gap allowance ($\eta(t_{mid})$) is as follows:
\begin{equation}
\phi(t_{mid}) = \frac{(t_{mid}-t_0)-\textrm{round}(t_{mid}-t_0)}{t_{dur}}
\end{equation}

\begin{equation}
\resizebox{.9 \columnwidth}{!}{$
\eta(t_{mid}) =
    \begin{cases}
           t_{cad} \times (16\phi(t_{mid})^{4}-8\phi(t_{mid})^{2}+2); &         \text{if  } -1\le\phi(t_{mid})\le1,\\
            \infty; &         \text{else}
    \end{cases}
    $}
\end{equation}
where $t_{mid}$ is the midpoint between the nearest neighbor measurements in the phase folded time series, $t_0$ is the ephemeris time, $t_{dur}$ is the fit transit duration, and $t_{cad}$ is the light curve cadence (nominally 0.0204 days). $\eta(t)$ provides an allowance of $t_{cad}$ during ingress and egress and a value of $2t_{cad}$ at the ephemeris. $\eta(t_{mid})$ quickly becomes very large for $t_{mid}$ values beyond $t_{dur}$ ($\eta(\phi=1,-1)=10t_{cad}$), but still requires some baseline coverage immediately before ingress and after egress. If $\Delta t_{mid} >\eta(t_{mid})$ (where $\Delta t_{mid}$ is the temporal spacing between the two measurements that define $t_{mid}$) for any value of $t_{mid}$, the TCE is flagged as a FP. This metric will flag TCEs with large gaps in the phase folded transit signal. As mentioned in Section \ref{sec:skye} the {\tt EVEREST} software masks many of the spacecraft operation cadences, but the neighboring cadences can occasionally produce outliers that line up to form TCEs. Such cases will produce gaps in the phase folded transit signal and be flagged as a FP by this metric.

Additionally, to ensure large gaps do not exist in the transit phase coverage, similar to that done in Section A.3.7.1 of \citet{tho18}, we implement a minimum 70\% in transit phase coverage requirement. Any TCEs that contain phase folded temporal gaps larger than $0.3t_{dur}$ will be flagged as a FP.         

\subsection{Period and Transit Duration Limits}
As mentioned in Section \ref{sec:har}, we used a harmonic fitter to remove stellar noise at periods less than 0.5 days. To avoid contamination caused by the introduction of this harmonic, we limit our sample to $P\ge0.5$ days. This limit has been enforced on {\tt TERRA} in Section \ref{sec:terra}, but such grid search algorithms can still provide detections beyond this limit by slightly mis-identifying the true transit period. Furthermore, TCEs found with periods very near 0.5 days can move below this limit when implementing the parameter search method of {\tt EDI-Vetter}. We strictly enforce this limit by again checking that $P\ge0.5$ days, otherwise the TCE is deemed a FP.

Another common FP is when a stellar oscillation has a short period ($<1$ day) falling in the forbidden detrending window (noted in Section \ref{sec:har}). In addition to our harmonic test (Section \ref{sec:har}), a symptom of a sine wave TCE is a very long $t_{dur}$ when compared to the signal period. Thus any TCEs with $t_{dur}/P>0.1$ are flagged as FPs. This also ensures that the planetary parameters remain physical. Any transit with $t_{dur}/P>0.1$ will be orbiting very near the surface of the star ($a_{pl}< 3R_\star$).

We acknowledge that ultra-short period planets \citep{san14, ada16} will certainly be rejected by the discussed cuts. However, all attempts made to capture these missed planets introduced a significant number of FPs into our candidate sample, negating our goal of high reliability. Again, readers concerned with USPs should perform their own fine-tuned searches. 

\subsection{Period Alias Check}
\label{sec:alias}
Occasionally the {\tt TERRA} grid search will find a transit that is some integer multiple of the true signal period. In such cases, it is very common that the signal is deemed a FP because of the significant ``secondary eclipse'', or the transit is poorly fit causing several other FP flags to be triggered. Alternatively, a signal with a period less than 0.5 days can be found with some larger multiple of the true period and contaminate the candidate population. 

To avoid such cases, we measure the Gaussian likelihood ($log\mathscr{L}=-\frac{1}{2}\chi^2$) of the transit model with the fit transit period ($P$), and compare that with the likelihood of the same model with four other possible periods: $P/2$, $P/3$, $2P$, and $3P$. We also scale the semi-major axis to stellar radius ratio ($a_{pl}$) parameter accordingly to ensure the transit duration is consistent among all tested likelihoods. To avoid edge cases where noisy data could falsely score one of the likelihoods higher, we give the initially fit period a slight advantage. If any of the alternative periods produce a likelihood greater than 1.05, the likelihood of the initially fit transit period, the signal is flagged as a period alias.

Any signal flagged as a period alias will be refit and revetted with the corresponding highest likelihood period. This cycle will occur at most three times (in most cases it only occured once), allowing for higher order integer multipliers to be tested. Within each cycle, the FP flag will be reset and the vetting metrics will be reprocessed at the new period. We then take the vetting as it stands in the final cycle.     

\subsection{Consistency Score}
\label{sec:consistency}
One cannot help but worry about the edge cases that could potentially slip through the cracks of our vetting metrics. To combat such cases, we implement a consistency score limit. To calculate the consistency score of our vetted candidates, we run each of them through the {\tt EDI-Vetter} fit and vetting pipeline 20 times. Given the statistical nature of the MCMC parameter estimation (and the limited number of burn-in and test steps), each run will process the light curve slightly different, potentially pushing the candidate into the FP bin. We only label a TCE as a candidate if it comes out of the vetting metric as a FP 25\% of the time or less (Consistency Score $\ge0.75$). This metric is comparable to the ``Disposition Score'' made available for \emph{Kepler} DR25 \citep{tho18}.

\section{CDPP} \label{sec:cdpp}
In \citet{chr12} the Combined Differential Photometric Precision (CDPP) metric was defined to create a quantitative measure of the expected stellar variability and systematic noise.  In other words, the CDPP tells us how strong a signal must be in order to overcome the noise present in the light curve. Having such a metric is essential to be able to analytically determine the expected MES value of a given transit (MES $\propto$ Transit Depth/CDPP). Here, we investigate the following transit durations: 1 hr, 1.5 hr, 2 hr, 2.5 hr, 3 hr, 4 hr, 5 hr, 6 hr, 7 hr, 8 hr, 9 hr, 10 hr.   

To calculate this value, we consider the processed light curve for each target (after {\tt EVEREST}, both GP fitters, and the harmonic fitter have detrended the light curve). We then inject a transit with depth equal to the standard deviation of the baseline light curve and a period which will reflect $N_{tr}=9$ transits within the light curve ($P\approx8.31$ days). For clarity, this post processing injection is independent of the pre-processing signal injections considered in Section \ref{sec:injections}. The injections discussed in the current section are only intended to measure the residual noise within each light curve. Using a short period with several transits allows us to quickly cover a larger portion of the light curve parameter space. It also allows us to remain robust to data gaps and systematic within the light curve. If one of the transits were to fall on a data gap, it should only decrease the measure of MES by 6\% for that run. Repeated iterations of this simulation will minimize this offset. The transit duration is tuned via the $a_{pl}$ parameter to match the CDPP duration of interest. Then the light curve is examined with {\tt TERRA}, at a period range corresponding to the injected signal, and the recovered MES value is used to measure the CDPP:

\begin{equation}
\text{CDPP}=\text{depth}/\text{MES}\times\sqrt{N_{tr}}
\end{equation}

where depth and CDPP are given in units of PPM. Resetting the light curve after each run, these injections are made 50 times with a randomly selected ephemeris at each draw, ensuring sufficient coverage of the time series. To summarize, nine transits are injected into each light curve and this is repeated 50 times, providing 450 injections in each light curve. The recorded CDPP value is then derived using the arithmetic mean of these 50 CDPP measurements. Using nine transits, as done here, allows us to statistically cover the entire phase space of the data for the CDPP duration of 4 hr with only 50 samples, dramatically speeding up our calculations. This process is then repeated for each of the 12 CDPP durations. A machine-readable version of our table of CDPP values is available \href{datafile1.txt}{online}.

\begin{figure*}[!ht]
\centering \hfill \includegraphics[height=7.5cm]{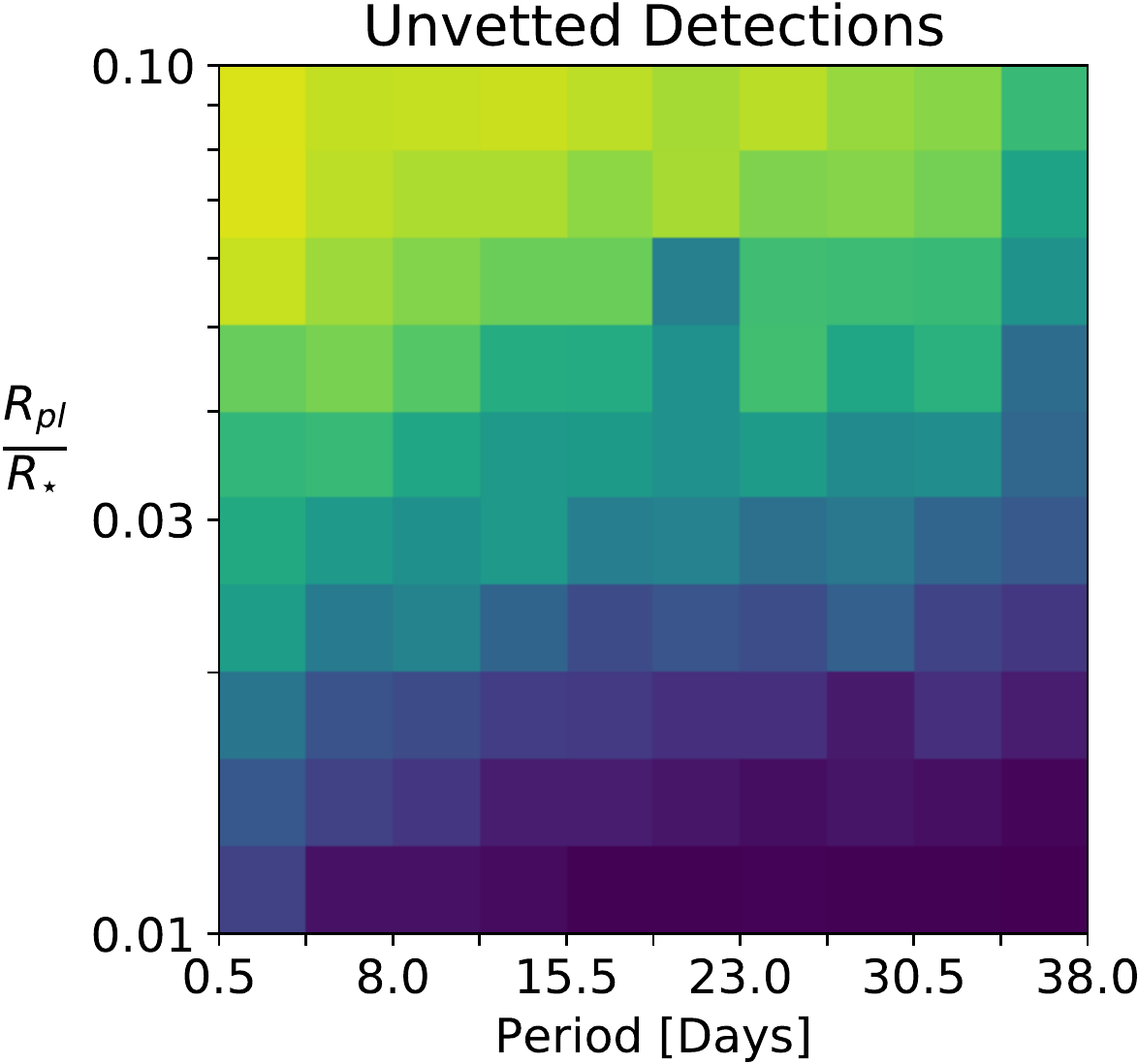}
\centering \hfill \includegraphics[height=7.5cm]{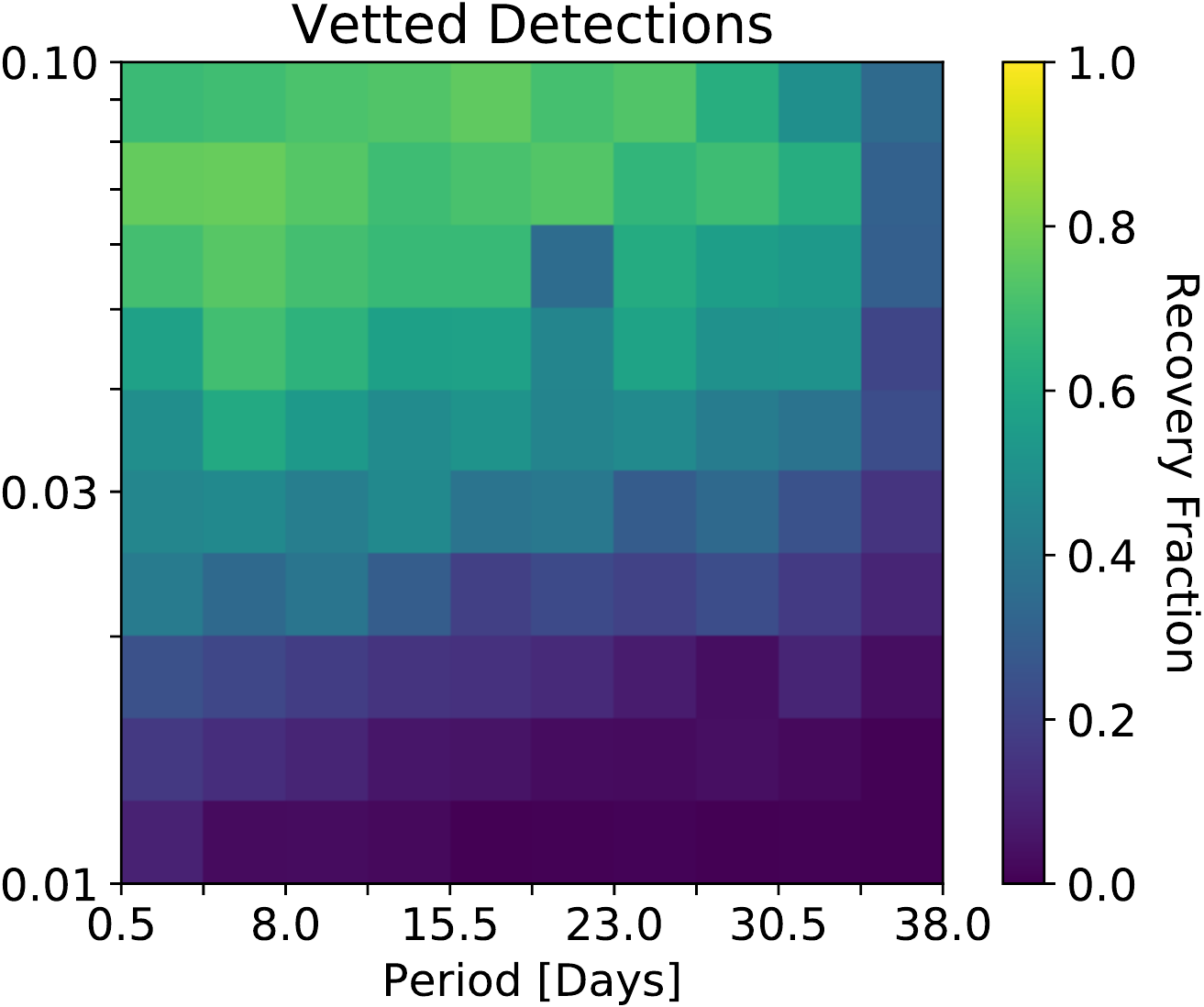}

\caption{The recovery fraction of injected planets as a function of period and radius. Each bin is sampled uniformly ($\sim250$ injections each). The \textbf{Left} plot shows the recovery faction for injected planets detected as a TCE. The \textbf{Right} plot shows the recovery fraction for injections that were detected as a TCE and also labeled a candidate by {\tt EDI-Vetter}. The anomaly seen at $P=20$ days and $R_{pl}/R_\star=0.05$, appears to be statistical fluctuation where an abnormally large fraction of high CDPP targets were selected for injection. We have thoroughly checked our pipeline and find no reason a signal in this specific range would have a unique recovery fraction. \label{fig:heatMap}}
\end{figure*}

\section{Injections} \label{sec:injections}

To test the likelihood of a signal being recovered, we inject artificial transit signals into the raw flux data. This is similar to the light-curve-level injections done by \citet{chr13,chr15,bur17}. By doing so, we can then run the flux measurements through the pipeline and count the number of recovered signals. This allows us to measure the detection efficiency of our pipeline.

Our injections utilize the {\tt batman} Python package to create our artificial transits \citep{kre15}. When injecting signals, it is important to maximize the number of signals near the MES threshold, this ensures good coverage of the detection efficiency curve \citep{chr13}. For each available light curve, we inject a transit signal which has a period uniformly drawn from a range of [0.5, 38] days and a $R_{pl}/R_{\star}$ drawn from a log-uniform distribution. The log-uniform draw for $R_{pl}/R_{\star}$ is meant to emphasize smaller planets where the signal is near the detection threshold. When injecting signals at the light-curve-level, it is important to be cautious about contaminating flux from nearby sources. If the neighboring source contributes a significant amount of flux to the target, the injections can be biased and the pipeline completeness will be artificially improved (smaller radius planets will appear easier to detect than what reality dictates). To combat this issue, we us the metrics described in Section \ref{sec:blend} to measure the amount of flux contamination. Instead of sampling from a range [0.01, 0.1] in log space for radius, we sample from [\resizebox{.9 \columnwidth}{!}{$0.01/\sqrt{\textrm{max}(F^{Tot}_{Gaia},F^{Tot}_{2MASS})}, 0.1/\sqrt{\textrm{max}(F^{Tot}_{Gaia},F^{Tot}_{2MASS})}$}] for each target. This ensures we do not bias our injection sample. In most cases, the flux contamination correction is negligible, so the original [0.01, 0.1] sampling space is recovered.   

It is important that these signals meet all of the vetting requirements to ensure we are not injecting planets that will certainly be rejected regardless of the detection capabilities of the pipeline. The most basic cut we make is on signals with less than three transits. We uniformly draw the ephemeris ($t_0$) from a range that will certainly produce three transits within the span of the data. This is uniform with a range of $[0, P]$ for short period injections ($<25$ days) and a range of $[0, t_{span}-2P]$ for longer period injections. Additionally, it is important that the signals do not trigger an eclipsing binary FP flag. In selecting the impact parameter ($b$) for the artificial planet, we uniformly draw from [0, 1], but such cases where $b$ and $R_{pl}/R_{\star}$ are large will result in a FP flag. We test our randomly drawn values against Equation \ref{eq:EB}, and if they exceed this limit, both values are redrawn. This cycle continues until the $b$ and $R_{pl}/R_{\star}$ values meet this criteria. Drawing impact parameters in this fashion leads to slight bias in our measure of completeness, as we force the injections to meet one of our detection metrics. However, this effect can be easily accounted for as discussed in Section \ref{sec:rec}. For $a_{pl}$ and the limb darkening parameters ($u_{1}$ and $u_{2}$), we calculate the expected values using the ATLAS model coefficients for the Kepler bandpasses tabulated by \citet{cla12} and the photometerically-derived stellar parameters available on \href{https://exofop.ipac.caltech.edu}{ExoFOP} \citep{hub16}. If such values are not available for the target in question we assume solar values. This assumption is made for less than 1\% of the targets and will only impact the inferred limb darkening parameters, avoiding the introduction of serious biases to our targets. The updated stellar parameters discussed in Section \ref{sec:stellar} were not available when this processing occurred. In testing, we found this use of older stellar parameters had no effect on our ability to recover signals. All final inferred planetary parameters were calculated using the most up-to-date stellar values (Hardegree-Ullman et al. submitted).

\begin{figure}
\centering \includegraphics[height=6.5cm]{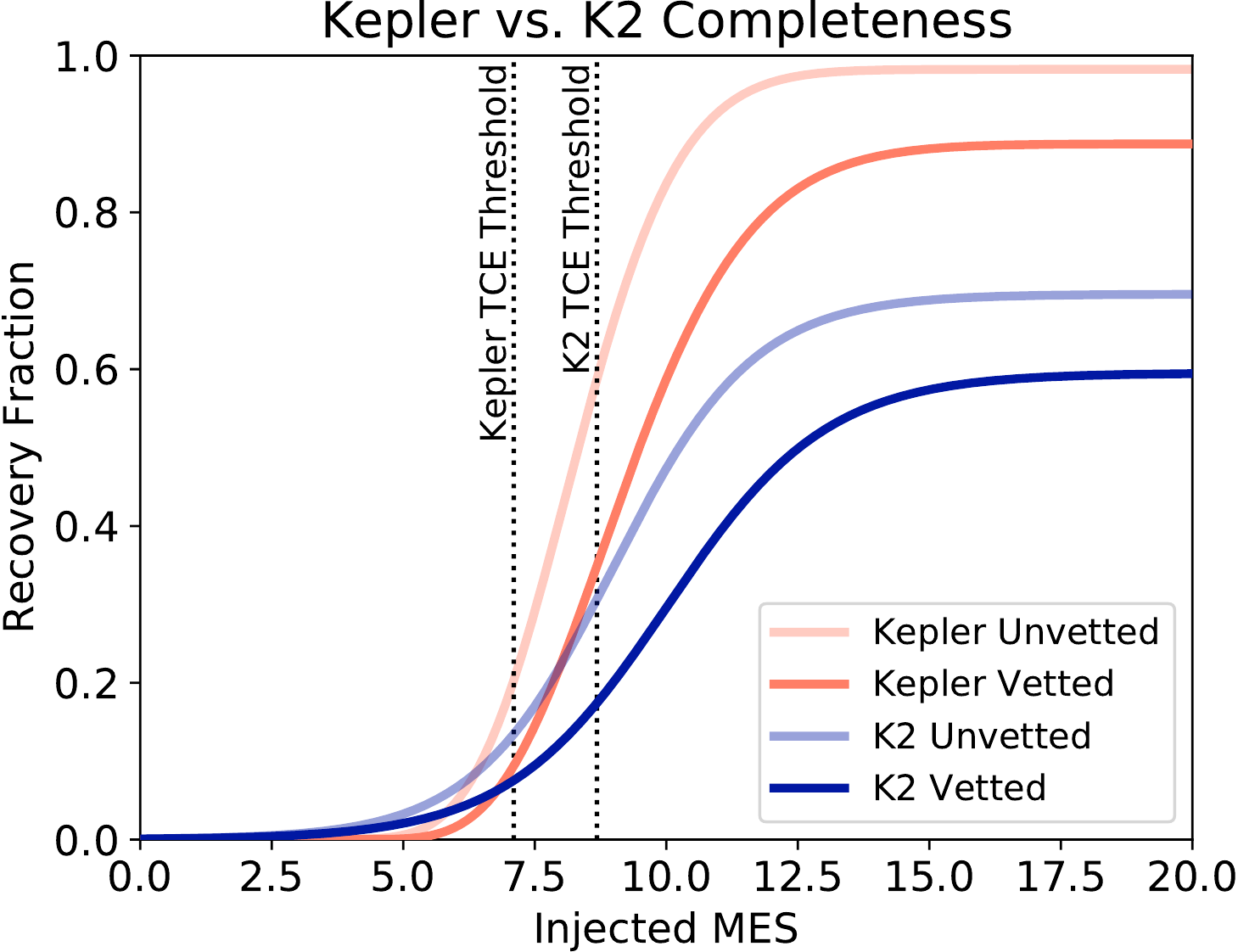}
\caption{A comparison of the completeness functions for \emph{Kepler} and \emph{K2}. The light peach and peach lines represent the Gamma functions for the \emph{Kepler} DR25 TCE \citep{chr17} and the {\tt Robovetter} \citep{cou17b} completeness respectively. The light blue and blue lines represent the logistic functions for the \emph{K2} TCE and {\tt EDI-Vetter} completeness functions derived in this study. All functions have been fit using injections across all period available ranges.  \label{fig:kepk2}}
\end{figure}

The injected raw flux data is then processed through our pipeline ({\tt EVEREST}, {\tt TERRA}, and {\tt EDI-Vetter}). We consider the recovery of a signal as a TCE and as a vetted candidate. Looking at this recovery fraction in terms of period and radius (Figure \ref{fig:heatMap}), we see that the vetting metrics in {\tt EDI-Vetter} do not induce a significant loss in the recovery fraction ($\sim10\%$; which are comparable to that seen by the {\tt Robovetter} \citep{tho18} for \emph{Kepler} as seen in Figure \ref{fig:kepk2}). A large portion of the vetting loss comes from edge cases ($P\sim0.5$ days, $P\sim38$ days, and $R_{pl}/R_{\star}\sim 0.10$), where statistical fluctuations can easily push the parameters beyond the threshold of our vetting limits. In general, we find that the pipeline has the most difficulty recovering transits with long periods and small $R_{pl}/R_\star$, and the pipeline is most adept at recovering signals with short periods and large $R_{pl}/R_\star$. This expected trend indicates that the detection efficiency of our pipeline is a function of MES. 

\begin{figure*}[!ht]
\centering \hfill \includegraphics[height=6.7cm]{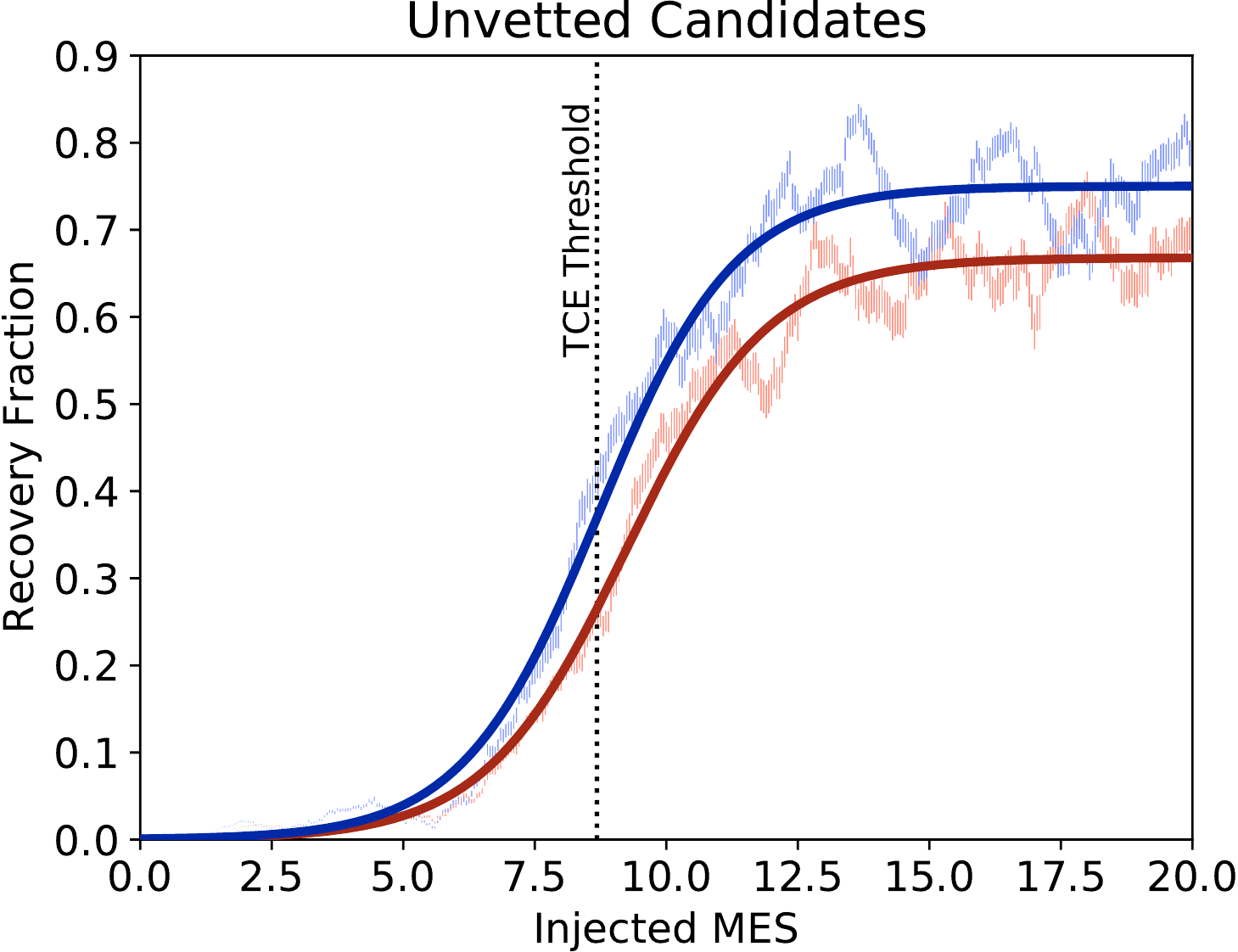}
\centering \hfill \includegraphics[height=6.7cm]{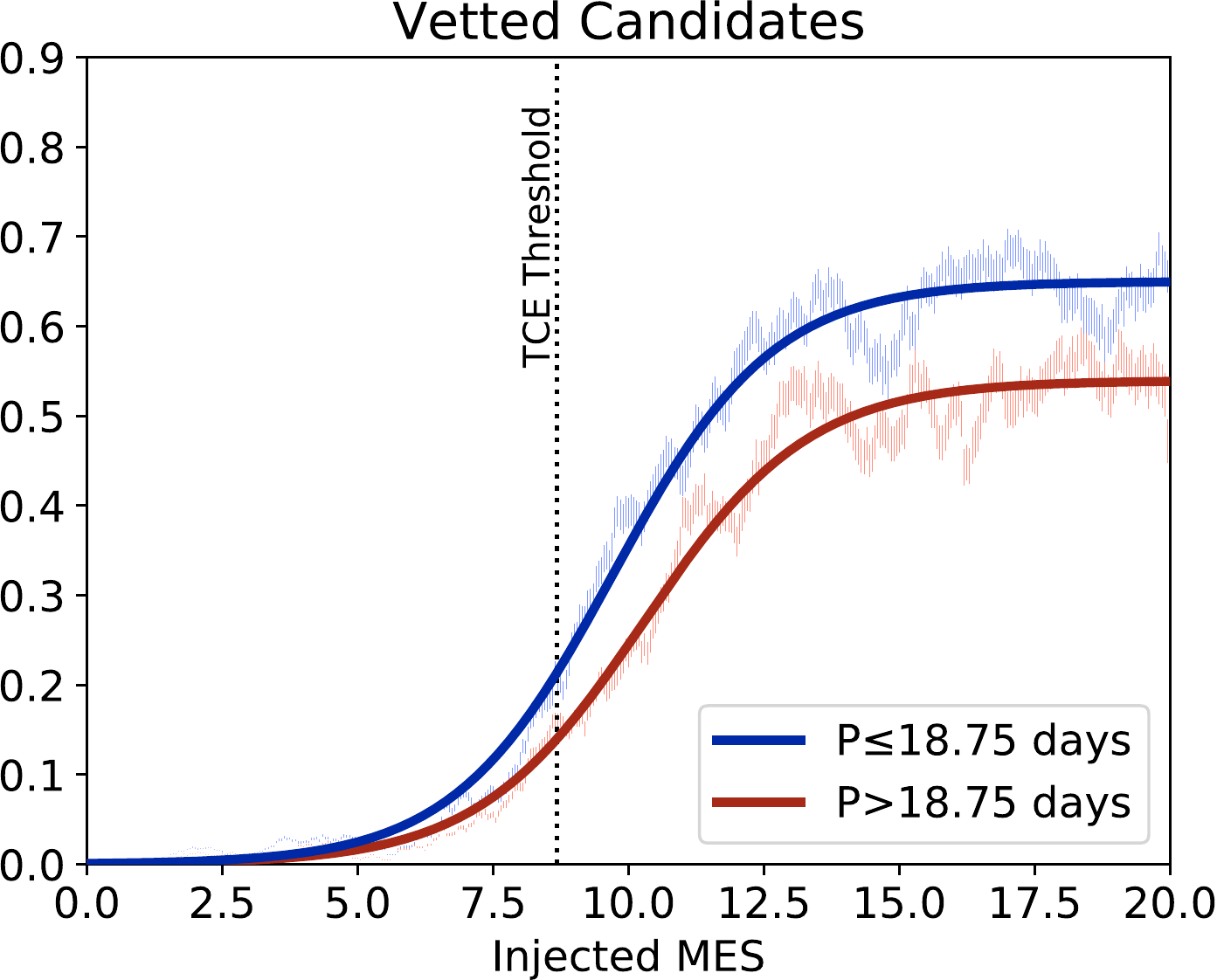}

\caption{Injected signal MES as a function of the recovery fraction. The \textbf{Left} plot shows the recovery function for planets detected as a TCE. The \textbf{Right} plot shows the recovery function for planets detected as a TCE and also deemed a candidate by {\tt EDI-Vetter}. The blue and red vertical lines represent the Kernel Density Estimation (KDE) using a uniform kernel with a width of 1 MES. The height of these lines represents the binomial uncertainty of each measurement. The dark blue and red lines illustrate the best fit logistic function given by the parameters in Table \ref{tab:MES}. We choose to separate the data at large periods to account for the decreased signal recovery from injections with only a few transits. All injections with periods $<18.75$ days will have at maximum three transits. \label{fig:MES}}
\end{figure*}

It is essential that we understand the noise (CDPP) within each light curve so that we can analytically determine the expected MES value for each injection. This will allow us to accurately calculate the recovery fraction throughout the MES range in question. We use the following equation to determine this parameter:
\begin{equation}
\text{MES}=\text{MES}_{Cor}\times\frac{\text{depth}}{\text{CDPP}_{t_{dur}}}\times\sqrt{N_{tr}}
\label{eq:MES}
\end{equation}
where $\text{CDPP}_{t_{dur}}$ is the CDPP value corresponding to the injected transit duration and $\text{MES}_{Cor}$ is the correction factor which matched the analytic function to the MES being detected by the pipeline ($\text{MES}_{Det}$). Since the CDPP values derived in Section \ref{sec:cdpp} are discrete, $\text{CDPP}_{t_{dur}}$ is an interpolation between the nearest duration values. Cases where $t_{dur}<1$ hr or $t_{dur}>10$ hr are not extrapolated, but rather assigned the nearest CDPP value. To calculate $\text{MES}_{Cor}$ we look at the most well behaved injections ($\text{MES}_{Det}<200$ and $b<0.1$). These restrictions limit the amount of MES smearing \citep{bur17} our injections will experience. This type of smearing occurs because the detected MES value is drawn from a distribution of values centered around the theoretical MES value. Larger MES and high impact parameter signals are prone to wider deviations from the theoretical MES value, making it more difficult to estimate the theoretical values. To minimize the potential bias this smearing effect can impose, we take the median value of $\text{MES}_{Det}/\text{MES}$, which we find to be 0.880, as our $\text{MES}_{Cor}$ value. This indicates that our analytic function will overestimate the MES value by 12\% without this correction factor. This factor is much more significant than the one found for \emph{Kepler} \citep[4.4\%;][]{chr17}, but is not surprising as the \emph{K2} data is more unruly. It is important to remember that even with this correction in place, the analytic MES value may differ from the $\text{MES}_{Det}$ on any given signal; the analytic MES value represents an average over the entirety of noise within the light curve, while $\text{MES}_{Det}$ is a discrete measurement.

\begin {table}
\caption {The logistic function parameters that best model the MES recovery fraction. These values were obtained by maximizing the binomial likelihood function for the KDE measurements provided in Figure \ref{fig:MES}.     \label{tab:MES}} 
\centering
\begin{tabular}{l c c c} 
\hline \hline \multicolumn{1}{l}{} & \multicolumn{1}{c}{$a$} & \multicolumn{1}{c}{$k$} & \multicolumn{1}{c}{$l$}\\
\hline 
            Unvetted &   &     &     \\
            $P\le18.75$ days & $0.75023$ & $0.7769$ & $8.7199$\\
            $P>18.75$ days & $0.6680$ & $0.7434$ & $9.2445$\\
            
            Vetted &   &     & \\
            $P\le18.75$ days & $0.6499$ & $0.6813$ & $9.7288$\\
            $P>18.75$ days & $0.5394$ & $0.6572$ & $10.2787$\\
           
\hline
\end{tabular}
\end{table}

As done in previous completeness measurements, we consider the separation of M dwarfs from the FGK stars. Typically M dwarfs have more correlated stellar noise and produce, on average, a decreased completeness function. We begin this separation with our initial list which contains 25,040 stellar targets. Removing targets with apertures that are too large or have incomplete data sets, we retain 24,949 targets. We then use the stellar parameters provided by Hardegree-Ullman et al. (submitted) to establish the spectral type classifications and $\log g$ values for our targets. To remove giants from our sample we eliminate targets with $\log g < 4$ (17,045 targets remain). We also remove targets that exhibit noisy light curves (CDPP$_{\rm{8 hr}}>1,200$ ppm). We select this limit of CDPP as it roughly corresponds to $5\sigma$ from the median CDPP value of all targets. This cut leaves us with 15,698 targets. We now separate the sample into FGK dwarfs using the spectral classification provided in Hardegree-Ullman et al. (13,046 targets) and M dwarfs (2,652 targets). We initially consider these two populations separately in our completeness calculation. However, we find no statistically significant difference in the detection efficiency between the two spectral groups. This is possibly caused by the small sample of M dwarfs available in Campaign 5, and we will continue to test this potential separation of stellar spectral type. 

Looking at bins of width 1 MES, we calculate the fraction of recovered injections as a function of MES. We use the analytic MES values found using Equation \ref{eq:MES} for both recovered and lost signals to ensure consistency. We also look to see what fraction is recovered once the signals have been processed through {\tt EDI-Vetter} (Figure \ref{fig:MES}). To account for inner bin fluctuations, and provide a more precise measure of recovery, we use the Kernel Density Estimator (KDE) for fitting. To calculate the KDE, we use a uniform kernel with a width of 1 MES. This method essentially recalculates the recovery rate at each shift of 0.05 MES. The uncertainties are calculated assuming binomial statistics, as the options are binary (the signal is either recovered or not recovered). The measured KDE values are then used to fit a logistic function ($f(x)$) to the data:
\begin{equation}
f(x)=\frac{a}{1+e^{-k(x-l)}}
\end{equation}
where $a$, $k$, and $l$ are all tunable parameters. In testing, we tried several functions, including the commonly used gamma function \citep{chr15,chr17}, but found that the logistic function provides the best fit with the detection efficiency of our pipeline. To fit the data we maximize the binomial log likelihood function ($\log L(x|\text{found},\text{lost})$):
\begin{equation}
\resizebox{.9 \columnwidth}{!}{$
\log L(x|\text{found},\text{lost})\propto(\text{found})\times\log(f(x))+(\text{lost})\times\log(1-f(x))
$}
\end{equation}
where `found' represents the number of recovered injections in a given KDE measurement and `lost' represents the number of lost signals in a given KDE measurement. The results of our fitting procedure are provided in Table \ref{tab:MES}.

It is apparent that the \emph{K2} pipeline is far less efficient than the previous \emph{Kepler} study. Comparing the unvetted candidates (70\% efficient at high MES) with the results of \citet{chr17} (94\% efficient at high MES), we can see that a 24\% drop in the recovery rate from \emph{Kepler} to \emph{K2} in Figure \ref{fig:kepk2}. This is significant, but expected as \emph{K2} data have far more systematic noise issues, causing transit detection to be more difficult. However, understanding this deficiency allows us to account for the loss when making occurrence measurements.         

We note that a more thorough test would be to inject the transits in at the pixel-level as done by \citet{chr17} for the \emph{Kepler} pipeline, however this process is far more computationally expensive and only provided a small correction (an increase of 2\%) to the \emph{Kepler} detection efficiency. We will consider such undertakings in future studies of the \emph{K2} completeness.

\section{Reliability} \label{sec:reliability}

To assess the reliability of catalog against instrumental false positives, we flip upside-down the processed light curves and run them through {\tt TERRA} and {\tt EDI-Vetter}. Since the light curve inversion ensures no true astrophysical transit exists within our data set, counting up the number of false signals that make it through the vetting process gives us a handle on the reliability of our candidates. Our pipeline finds 6,145 TCEs in the inverted light curves (compared to 5,452 found from the original processing), and only four (0.07\%) of these make it through the {\tt EDI-Vetter} testing.

\begin{figure}
\centering \includegraphics[height=6.0cm]{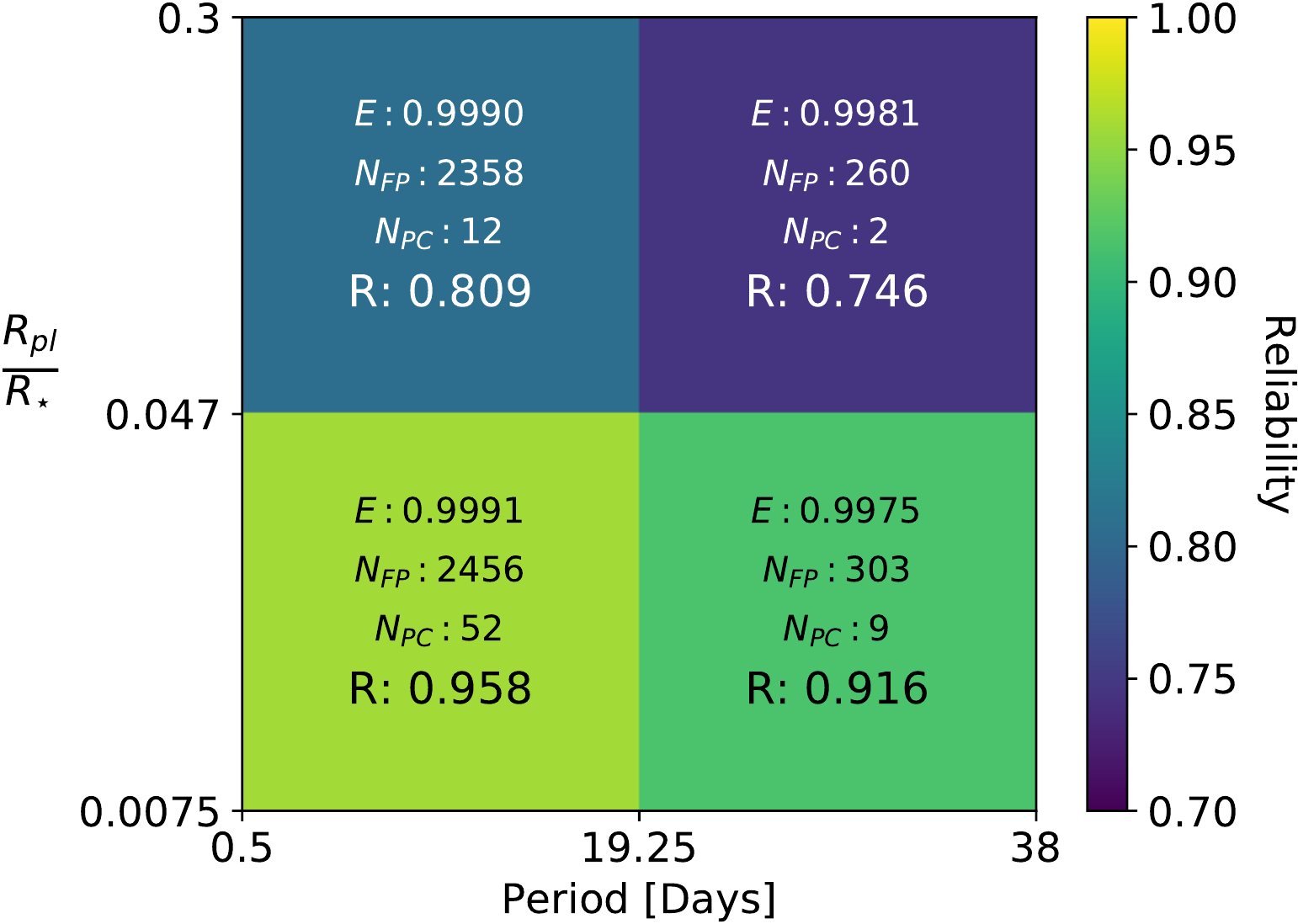}
\caption{A display of the reliability of our pipeline. The $N_{FP}$ values correspond to the number of FPs identified in each bin. The $N_{PC}$ values indicate the number of planet candidates found in each bin. The $R$ value is the expected reliability given the number of candidates and FPs using Equation \ref{eq:reli}. The bins have been divided at the mid-point of the ranges considered in this search. The $R_{pl}/R_{\star}$ cut is at the mid-point in log space. \label{fig:reli}}
\end{figure}

To determine how reliable our pipeline is, we use the metrics laid out in Equations 3-8 of \citep{tho18}. First, we must calculate how efficiently FPs are being classified by the pipeline ($E$), using the inverted light curve results:
\begin{equation}
E\approx\frac{N_{FP_{\mathrm{inv}}}}{T_{FP_{\mathrm{inv}}}}
\end{equation}
where $T_{FP_{\mathrm{inv}}}$ is the total number of TCEs found in the inverted light curves and $N_{FP_{\mathrm{inv}}}$ is the number of TCEs found that were vetted and accurately assigned FP status. Since our inverted light curve test only produced four falsely identified candidates, small number statistics need to be considered. To account for these small number statistics, we add one additional false alarm signal to each measure of $E$ so that it becomes:   
\begin{equation}
\label{eq:E}
E\approx\frac{N_{FP_{\mathrm{inv}}}}{T_{FP_{\mathrm{inv}}}+1}
\end{equation}
This addition is motivated by the expectation value for $\lambda$ given $N$ measurements from a Poisson distribution using Bayes theorem:
\begin{equation}
\resizebox{.9 \columnwidth}{!}{$
\begin{aligned}[b]
P(\lambda | N) & \propto P(N|\lambda)*P(\lambda)\propto\lambda^N e^{-\lambda} * 1\propto \text{Gamma}(shape=N+1,scale=1)\\
\big<P(\lambda | N)\big> & =\big<\text{Gamma}(N+1,1)\big> =N+1
\end{aligned}
$}
\end{equation}
where Gamma represents a Gamma distribution. However, this means we are likely underestimating the efficiency of our FP identification. Thus, our reliability values should be considered a lower limit that will likely increase as we increase our FP count with data from other \emph{K2} Campaigns. Overall, our pipeline finds $N_{FP_{\mathrm{inv}}}=6,141$ and $T_{FP_{\mathrm{inv}}}=6,145$, using Equation \ref{eq:E} our overall pipeline $E$ is 0.9992. In Figure \ref{fig:reli} we report the $E$ values calculated for each region of parameter space.  

Using the number of planet candidates ($N_{PC}$) and the number TCEs given FP status ($N_{FP}$) in the non-inverted light curves, we can determine the reliability fraction ($R$) of our data set:
\begin{equation}
R=1-\frac{N_{FP}}{N_{PC}}\Bigg(\frac{1-E}{E}\Bigg)
\label{eq:reli}
\end{equation}
We provide a coarse binning of these reliability metrics in Figure \ref{fig:reli}. Overall our pipeline finds a reliability fraction of 0.942, indicating that we should only expect $\approx4$ FPs to be contaminating our candidate catalog. Again, small number statistics are largely at play here, and it is very possible our true reliability is even greater than the values we have reported. This is slightly less reliable than the \emph{Kepler} pipeline, which finds an overall reliability of 0.97 for their all stars measure \citep{tho18}. However, a more appropriate comparison would consider the parameter space where a similar number of transits occur (Periods $>10$ days for \emph{Kepler}). Using Figure 8 of \citet{tho18} and knowing that 18,660 TCEs were detected with periods beyond 10 days, we estimate this reliability to be roughly 0.95 for the \emph{Kepler} data set in a comparable parameter space. While this value is still higher than the measurement presented by our pipeline, we remind the reader that a conservative estimate of reliability has been provided which will likely improve as we increase our sample.      

\begin{figure}
\centering \includegraphics[height=6.2cm]{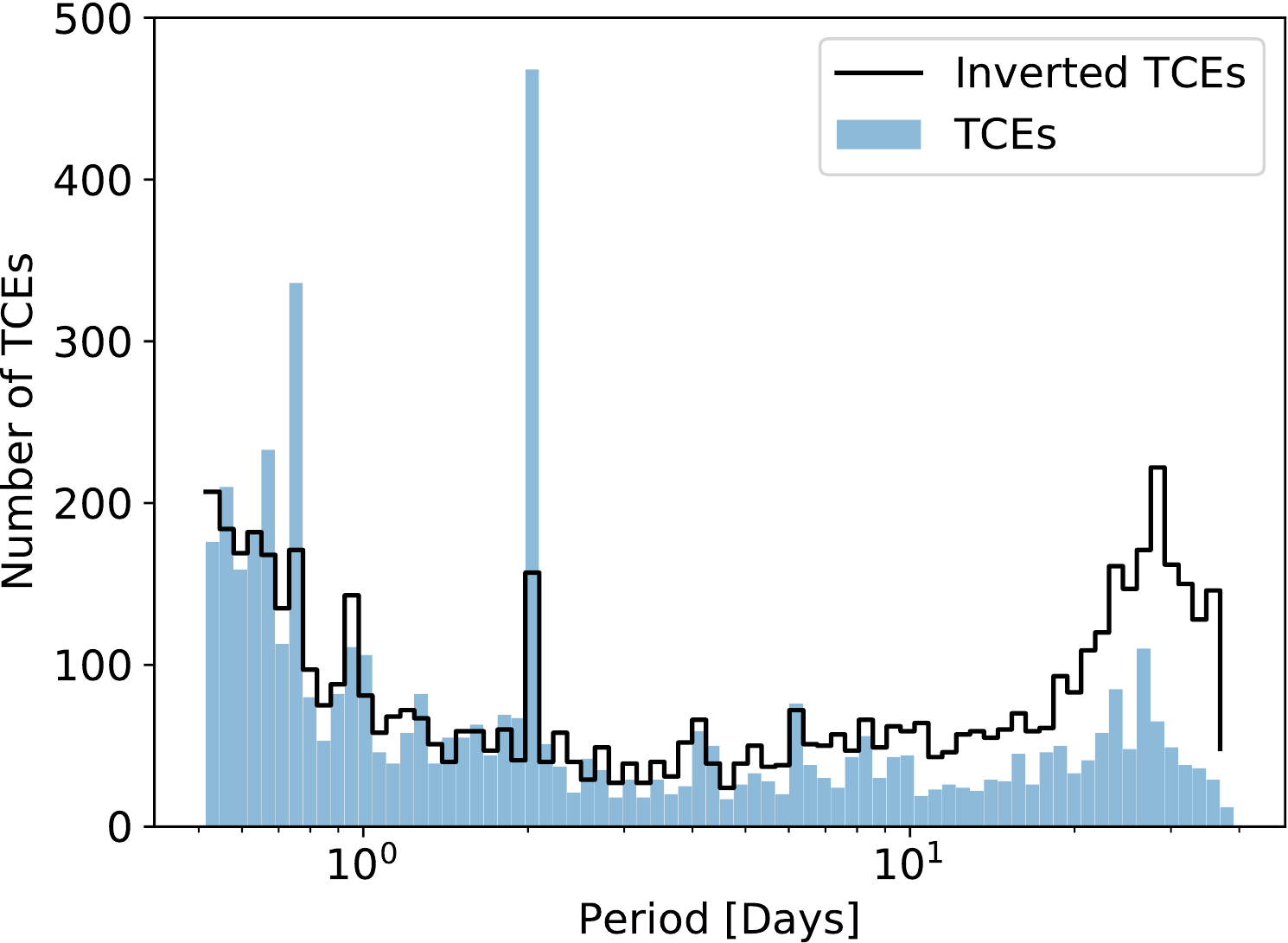}
\caption{ A plot of the TCE distribution and the inverted light curve TCE distribution.  \label{fig:reli2}}
\end{figure}

Testing reliability by inverting the light curves assumes the false positives are symmetric and will persist upon inversion. One method of testing this assumption is by examining the period distribution of the TCEs. In Figure \ref{fig:reli2} we compare the TCE distribution to the inverted TCE distribution. Overall, the distributions provide a reasonable match. We find some under representation in the inverted TCEs near the Resat period (2 days) and the third thruster harmonic (0.75 days), indicating these systematics are slightly asymmetric. Additionally, the long period TCEs are slightly over represented in the inverted light curves. We acknowledge that these mis-matches could modify our measure of reliability. Additional light curves from other campaigns will help improve this measure and other methods such a data scrambling (see Section 2.3.2 of \cite{tho18}) may be considered in future studies.

\section{Window Function}
\label{sec:window}
The window function gives the probability ($prob$) that a certain period ($P$) will meet the three transit minimum requirement for TCE consideration within the span ($t_{\text{span}}$) of the available light curve \citep{bra09,bal10}. This metric is an essential ingredient for occurrence measurements \citep{bur15,zin19,bry19}, where all aspects of detection probability must be considered.  If the data were seamless, without any masking, the equation can be found analytically:
\begin{equation}
\label{eq:win}
\begin{aligned}[t]
prob & =1; & P<t_{\text{span}}/3\\
prob & =\frac{t_{\text{span}}}{P}-2;  & \frac{t_{\text{span}}}{3}\le P \le \frac{t_{\text{span}}}{2}\\
prob & =0;  & P> t_{\text{span}}/2
\end{aligned}
\end{equation}

where the $t_{\text{span}}$ for Campaign 5 is nominally 74.82 days. However, this problem becomes more difficult to solve when considering masked cadences within each light curve. While some cadences are masked globally (see Section \ref{sec:skye}), often the masking is done on an individual per-target basis, in an attempt to remove outliers. To ensure accuracy, we directly measure this probability for each target light curve, similarly to \citet{bur17b} for \emph{Kepler} DR25. 

\begin{figure}
\centering \includegraphics[height=6.7cm]{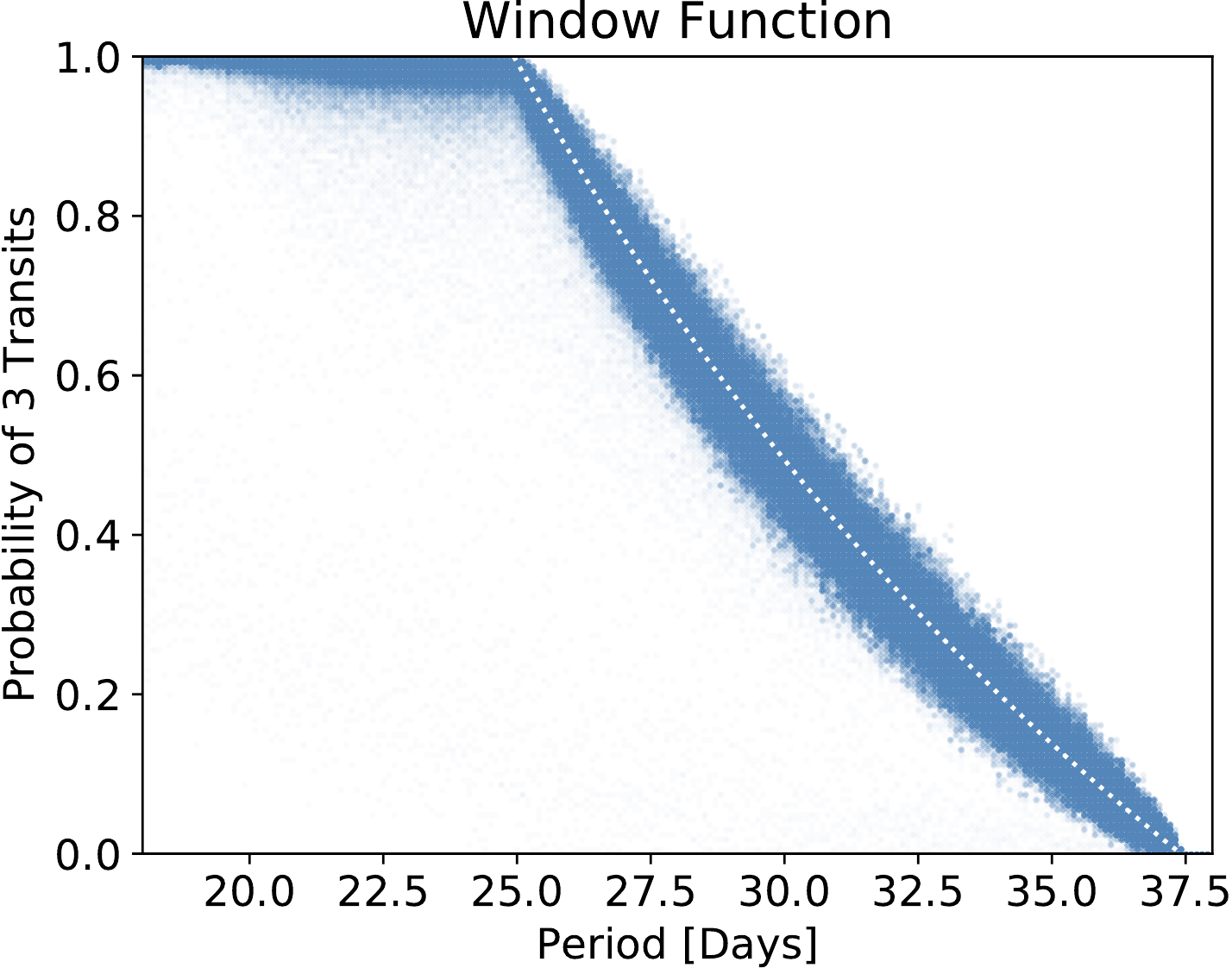}
\caption{The measured window function for all Campaign 5 targets. The white dotted line represents the analytic function (Equation \ref{eq:win}) expected for this campaign, assuming $t_{\text{span}}=74.84$ days. We see that a large fraction of the targets appear to exceed the functional probability, which represents the best case detection scenario. This is due to the increased $t_{\text{span}}$ for a limited number of targets, but more importantly, from statistical fluctuations in our Monte Carlo simulation. The increased dispersion seen near $prob=0.5$ is the signature of a binomial distribution, caused by statistical fluctuations.        \label{fig:win}}
\end{figure}

Looking at each light curve, we used the corresponding masked cadences (those which were masked prior to the TCE search process) to create the gaps in our light curve. Starting at 18 days, we take steps of 0.1 days and test the probability of each period yielding three transits. It is important here to consider how the transit duration will affect the number of available cadences in each transit. Longer duration transits will have a higher likelihood of avoiding serious masking. To marginalize over this parameter, we sample from our planet candidate empirical sample with replacement. This sampling is biased toward planets with longer transits (which are easier to detect). To account for this, we do not sample uniformly, but rather with a $1/\sqrt{t_{\text{dur}}}$ probability (as MES scales with $\sqrt{t_{\text{dur}}}$). This gives more weight to shorter transit durations. This weighted sampling produces a median $t_{\text{dur}}$ of 2.20 hours with a minimum of 1.05 hours, and a maximum of 8.38 hours. A more thorough metric would be to create a 2D grid, stepping through both period and $t_{\text{dur}}$ values, but we found little variation when changing the transit duration, and have settled on a simple measure using only the planet period. At each step of period, we sample 160 times (roughly 2\% accuracy for a binomial distribution), selecting a new $t_{\text{dur}}$ and emphemeris ($t_0$) at each iteration. We then count up the number of times three or more transits occurred within the span of the data set. We only consider a transit observed if at least 50\% of the transit was unmasked. We then record the probability of three transits occurring at this period and move on to the next step in period space. Figure \ref{fig:win} shows all of the measured window functions. It is apparent that the masking is spread uniformly though the light curves, minimizing the deviation from the analytic equation (\ref{eq:win}). If significant gaps exist in the light curves, the measurements produce large deviations from the functional form. To first order approximation, the analytic equation provides a good measure of the window function. For a more detailed calculation, we provide the measured window function for each target in the \href{datafile2.txt}{online} version of this article.

\begin{figure}
\centering \includegraphics[width=\columnwidth]{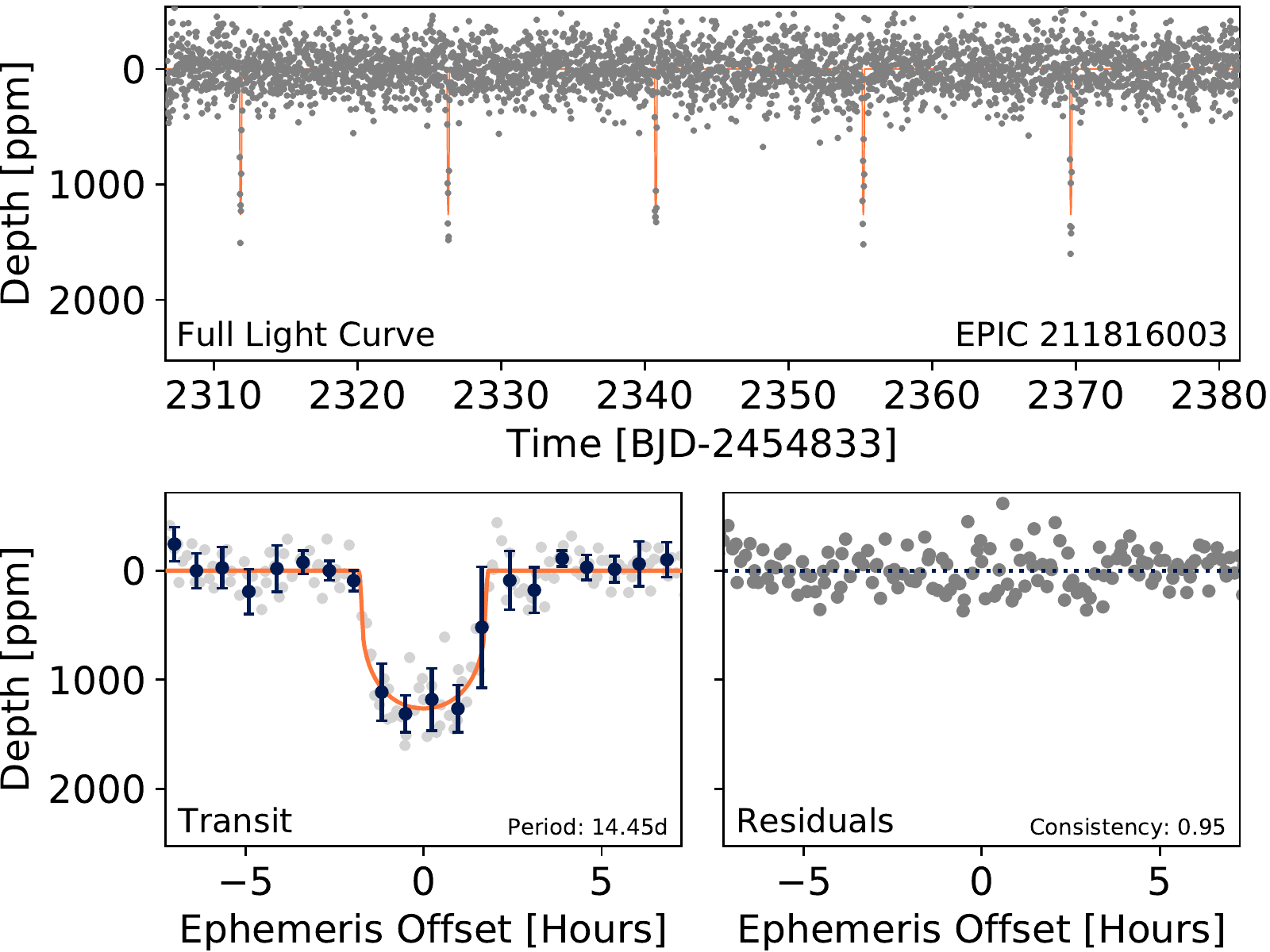}
\caption{An example of a fully vetted planet candidate. The grey points represent the processed flux measurements and the navy points show the binned average. Here we choose a bin width that ensures five bins exist within the transit duration. The orange line represents the best fit transit model.   \label{fig:exm1}}
\end{figure}

\section{Candidates} \label{sec:candidates}

Using our described pipeline, we detected 75 planet candidates in the 15,698 C5 light curves searched. The parameter estimations made by the pipeline are performed with a focus on computational speed. With our final list of 75 candidates, we can afford to let our MCMC more thoroughly explore the likelihood surface. We use {\tt emcee} \citep{for13} with 100 semi-independent walkers, a 500 step burn-in, and a 500 step parameter space coverage (50,000 test samples). In this fit we allow period ($P$), impact parameter ($b$), emphemeris ($t_0$; with a zero offset of 2454833 days), radius ratio ($R_{pl}/R_{\star}$), and semi-major axis ($a$) to vary at each iteration. We assume circular orbits for all of our planets and use the limb darkening parameters derived from the existing stellar parameters (we discuss these non-variable transit parameters in Section \ref{sec:injections}). Table \ref{tab:Cand} shows an example version of our results. We provide a full machine-readable version of this data \href{datafile3.txt}{online}. For each candidate, we also make a snapshot of each transit, which includes the full light curve, the folded transit, and the residuals when the transit model is removed. We provide an example of this snapshot for planet candidate EPIC 211816003.01 in Figure \ref{fig:exm1}. We provide our full list of candidate snapshots on \href{https://exofop.ipac.caltech.edu}{ExoFOP}.

As mentioned in Section \ref{sec:injections}, possible flux contamination can cause an underestimation of the transit depth. We use the parameters described in Equation \ref{eq:EB} to correct for detected contamination. All $R_{pl}/R_{\star}$ parameters are multiplied by the flux contamination factor $\sqrt{\textrm{max}(F^{Tot}_{Gaia},F^{Tot}_{2MASS})}$ increasing the ratio accordingly. We note that this factor only accounts for flux contamination detected by either Gaia or 2MASS. It is likely that other transit dilution exists from bound binary companions \citep{cia17,fur17,mat18}. High resolution follow-up is essential to rule out such cases for our candidate list. Additionally, signal processing has the potential of artificially decreasing the transit depth. We discuss this issue in greater detail in Section \ref{sec:plRad}. We note here that our listed candidates do not reflect any type of systematic correction, as we leave such a procedure to be carried out at the reader's discretion.

Within our candidate sample we find seven systems with more than one detected planet. System multiplicity is often a good indicator that the planets are real, and not FPs \citep{lis14}, thus we focus our discussion on these candidates. One target, EPIC 211428897, contains four planet candidates. Three of the candidates were previously noted \citep{dre17,pet18}. Alongside \citet{kru19}, we find an additional candidate at period of 6.265 days. We also verify the claim of \citet{kru19} for two new multi-planet systems: EPIC 211413752 and EPIC 212072539 both had a single known candidate \citep{pet18}, but we were able to extract additional planets in both of these systems, increasing the number of known multi-planet systems. In Section \ref{sec:stellar} we briefly discuss the how our multi-planet candidates are distributed among stellar spectral type. 

We suspect that several of the single planet systems likely have additional planets which either do not transit, or were not detected because of the detection order issues pointed out by \citet{zin19} for \emph{Kepler}. Many of these same detection features exist in this pipeline as discussed in Section \ref{sec:terra}. An example of this is EPIC 212164470 \citep{bar16,pop16,pet18,may18}, where we only recovered the transit with a period of 7.81 days, and did not recover the smaller MES signal at 1.74 days \citep{may18}. In later iterations of this pipeline, we will test the detection efficiency of multiplicity by injection of several planets into each light curve, allowing us to directly measure the role detection order plays in transit recovery. However, such a task is beyond the scope of this paper. 

One reason we lose multiple planet signals is due to the limited range of our period search. We limit our pipeline to a range of 0.5 days to 38 days. Previous surveys have found two systems that have multiple transiting planets with periods less than 0.5 days (EPIC 211305568; \citealt{dre17} and EPIC 211562654;	\citealt{may18}). Upon visual inspection of the light curve, it appears these candidates were found with our TCE search, but then rejected by our $\ge 0.5$ days vetting requirement. Weakening this vetting requirement may yield more planets, but we found in practise it will dilute the purity of our sample (Section \ref{sec:terra}). We note that a large fraction of the Campaign 5 targets are also available in either Campaign 16 or 18. By stitching the data from multiple campaigns together, the available data span can be extended to potentially find periods beyond our 38 day limit. An example of this is EPIC 211611158, which has a known planet at a period of 52.71 days \citep{may18}. It should be noted that this planet was not found by stitching, but rather by relaxing the three transit requirement. Currently, there has been no effort to combine overlapping campaigns in the manner discussed. We hope to develop the ability to stitch together campaigns in future iterations of this pipeline.

\begin{figure}
\centering \includegraphics[width=\columnwidth]{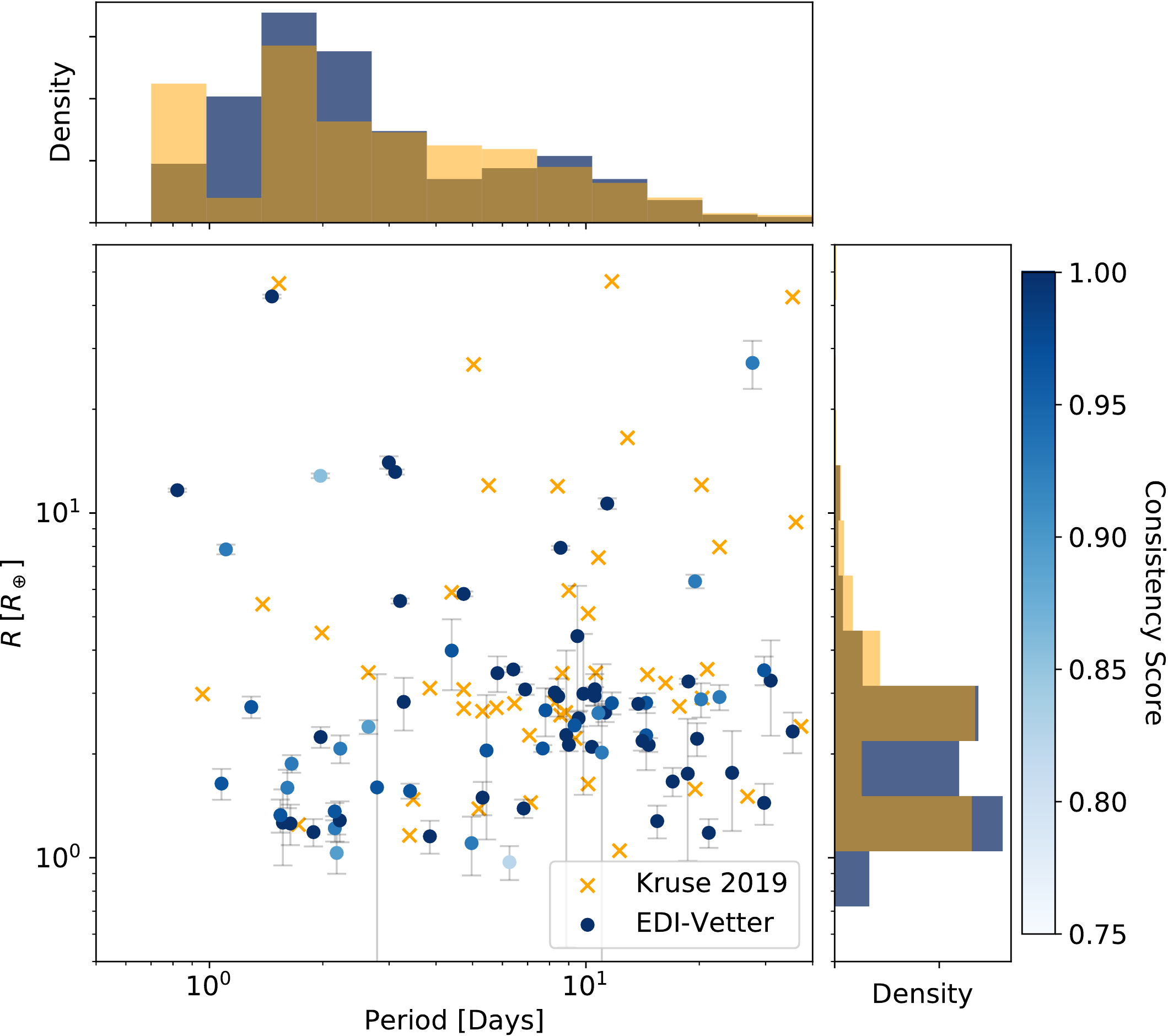}
\caption{A plot of the candidates resolved by our pipeline. The dots ({\tt EDI-Vetter}) correspond to the candidates detected, with a color scaling to show the consistency score of each candidate (Section \ref{sec:consistency}). The X symbols (Kruse 19) represent the candidates not found by our pipeline, but noted by \citet{kru19}. The histograms show the density of the found planets for both period and radius by our pipeline in blue and the entire \citet{kru19}  Campaign 5 catalog in orange. \label{fig:cand}}
\end{figure}

\begin {table*}
\caption {An example of the Campaign 5 planet candidate table available for download \href{datafile3.txt}{online}. Listed below are the new planet candidates detected in this paper.    \label{tab:Cand}} 
\centering
\begin{tabular}{l c c c c c c c} 
\hline \hline \multicolumn{1}{l}{Candidate} & \multicolumn{1}{c}{$P$} & \multicolumn{1}{c}{$R/R_{\star}$} & \multicolumn{1}{c}{$R$} & \multicolumn{1}{c}{$t_0$} & \multicolumn{1}{c}{$b$} & \multicolumn{1}{c}{$t_{dur}$} & \multicolumn{1}{c}{...}\\
\hline 

\multicolumn{1}{c}{} & \multicolumn{1}{c}{(days)} & \multicolumn{1}{c}{}  & \multicolumn{1}{c}{($R_\Earth$)}  & \multicolumn{1}{c}{(BJD)}& \multicolumn{1}{c}{} & \multicolumn{1}{c}{(hours)} & \multicolumn{1}{c}{...}\\
\hline 

$211562654.04$ & $2.15290\pm 0.00015$ & $0.01151\pm0.00062$ & $1.23\pm0.11$ & $2307.4643\pm0.0035$ & $0.47\pm0.14$ & $2.03\pm0.21$\\
$211711685.01$ &$15.4624\pm0.0023$ & $0.01273\pm0.00050$ & $1.28\pm0.06$ & $2310.8848\pm0.0054$ & $0.47\pm0.17$ & $3.94\pm 0.55$ \\ 
$211918985.01$ & $9.4892\pm0.0040$ & $0.0390\pm0.0018$ & $4.40\pm4.35$ & $2316.0782\pm0.0078$ & $0.26\pm0.19$ & $8.38\pm1.06$ \\
$211942755.01$ & $19.7288\pm0.0034$ & $0.0285\pm0.0013$ & $2.22\pm0.13$ & $2322.8335\pm0.0038$ & $0.688\pm0.056$ & $2.70\pm0.21$ \\
$211953244.01$ & $29.7236\pm0.0051$ & $0.01547\pm0.00097$ & $1.44\pm0.10$ & $ 2307.3432\pm0.0056$ & $0.67\pm0.18$ & $4.07 \pm 0.90$ \\
$211958340.01$ & $1.465460\pm.000028$ & $0.2693\pm0.0074$ & $42.49\pm1.72$ & $2307.17786\pm0.00082$ & $0.468\pm0.058$ & $3.23\pm0.19$ \\
$212020330.01$ & $1.65348\pm0.00039$ & $0.01430\pm0.00059$ & $1.87\pm0.14$ & $2307.8556\pm0.0051$ & $0.78\pm0.12$ & $2.69\pm1.20$ \\
$212119244.01$ & $1.07692\pm0.00023$ & $0.00772\pm0.00045$ & $1.64\pm1.62$ & $2307.1496\pm0.0055$ & $0.41\pm0.25$ & $2.02\pm0.47$ \\

\hline

\end{tabular}
\end{table*}

One system of known planets was completely lost by our pipeline, EPIC 212157262. This system was detected and also confirmed by \citet{may18}, who used the {\tt K2SFF} \citep{van14} processed light curves. Some of the planets in this system produce signals near the detection threshold, and were not detected in our pipeline TCE search. Using different detrending algorithms can yield slightly different results, which may help push some signals above this threshold while moving others below it. Using all the different detrending algorithms simultaneously can help minimize such loss \citep{kos19}, but requires a serious computational cost to create the completeness and reliability metrics available here. Thus, we have only used the {\tt EVEREST} detrending algorithm, and accept the potential planet loss. Additionally, two of the planets \citep[\emph{K2}-187 d,e;][]{may18} have strong signals and should not be near the detection threshold. In fact, these two planets were detected in our injection simulation. This loss is caused by the indeterministic nature of the {\tt EVEREST} software, meaning the same flux data can produce two different processed light curves on different iterations. One way to combat this problem would be to process all of the light curves twice and take the lowest CDPP case, but such a task would significantly increase the amount of computational time required, and is not done here. We run each light curve through {\tt EVEREST} twice, in doing so we acknowledge that an individual injected light curves may produce better or worse CDPP than our nominal observational light curves. However, when considering the overall injection population this effect should cancel out. Such issues are folded into our completeness measure in Section \ref{sec:injections}. Future iterations of our pipeline will attempt to minimize this effect by forcing {\tt EVEREST} to abide by a fixed seed.

 Our search pipeline is able to identify nine previously unknown candidates (see Table \ref{tab:Cand}). These new candidates are mostly small radius planets ($R_{\Earth}<3$) which were not detected by previous searches that used higher MES thresholds. One of these candidates extends the multiplicity of known systems (EPIC 211562654). Almost all of the new candidates have \emph{Kepler} band magnitudes $<13$, with the exception being EPIC 211918985.01 and EPIC 211942755.01 which are 16.4 and 14.1 magnitudes respectively. The brightness of these new candidates makes them excellent targets for follow-up validation. 
 
One candidates (EPIC 211958340.01) appear to be suspiciously large for planets candidates, highlighting the limitations of our stellar agnostic vetting metric (Equation \ref{eq:EB}). Both of these candidates orbit stars with large stellar radii ($>1R_\Sun$), where eclipsing binaries have the potential to produce small dips in the light curve. We suggest occurrence study make an upper limit radius cut to avoid these potential astrophysical false positives. Similar contaminates can be found in the \emph{Kepler} pipeline candidate list.                

Overall, our pipeline detected 75 planet candidates. Comparing this to known candidates from \citet{kru19}, who manually vetted transits signals, we recover 51\% of the known planet candidates within our period and radius ratio search limits. This manually vetted catalog provides the largest yield \emph{K2} candidate catalog to date and also used the {\tt EVEREST} software to remove spacecraft systematics, thus we use \citet{kru19} for comparison in Figure \ref{fig:cand}. It is important to remember that while \citet{kru19} may provide a more complete list of the Campaign 5 candidates, our catalog of candidates has been uniformly vetted, eliminating potential confirmation biases. Comparing to the systems of multiple planets, where FPs are less likely, we recover 60\% of the known multi-planet system candidates. An alternative comparison can be made for \citet{pet18}, who also used {\tt TERRA} to search for transit signals in Campaign 5 of \emph{k2}. We recover 70\% of the planets in the manually vetted \citet{pet18} catalog that fall within our period and radius ratio search limits. Two other large transit searches have been performed on Campaign 5, \citep{dres17} and \citet{pop16}. We find our candidates list overlaps with 51\% of each of these previous studies within our period and radius ratio search limits.

There are several reasons we fail to recover 100\% of the known planet candidates. As mentioned previously, using different detrending algorithms will yield different results, and cause us to lose some planets that a different algorithm would otherwise allow us to detect; 37\% of the lost planets were never detected as a TCE by our pipeline. Additionally, our vetting metrics are more harsh than some studies, which take an ``innocent until proven guilty'' philosophy toward planet candidacy. We aim for reliability, so we are willing to remove real signals at the benefit of ensuring fewer FPs. The Eclipsing Binary metric (Equation \ref{eq:EB}) of our vetting removes many of the potential candidates found by other studies, accounting for 17\% of the lost planets. The difference in even and odd transit depths may cause a rejection when one of the transits is poorly detrended (Section \ref{sec:evenodd}), causing 13\% of the known planet candidate rejections. The outlier metric (Section \ref{sec:out}), meant to remove signals dominated by the chance alignment of several outliers, can misclassify noisy transit signals, removing 13\% of the lost planets. The remaining 20\% are lost due to single triggers of various other metrics in {\tt EDI-Vetter}.

\subsection{Stellar Parameters}
\label{sec:stellar}

The planet radius measurements available in Table \ref{tab:Cand} were calculated using the stellar radius values given by Hardegree-Ullman et al. (submitted). These parameters were inferred using photometric \emph{g, r, J, H, K, Kepler,} and \emph{Gaia} band data in combination with the Large Sky Area Multi-Object Fibre Spectroscopic Telescope \citep[LAMOST;][]{su04} spectra. Using the well defined parameters (spectral type, $T_{\mathrm{eff}}$, $\log\,g$, [Fe/H]) available from the LAMOST data, a random forest algorithm \citep{ped11} was used to assign these parameters to each of the \emph{K2} tagets based on the photometry for targets without spectroscopic measurements. Bolometric luminosities were computed using \emph{K} band magnitudes and \emph{Gaia} parallaxes, from which stellar radii were calculated using the Stefan-Boltzmann law. In Table \ref{tab:Cand} we provide the transit parameters in such a way that stellar and planet parameters can be updated as new information becomes available.
Using this updated stellar classification, our list of candidates includes 58 planets around FGK stars and 17 planets that are hosted by M dwarf stars. We find 8 planets in multi-planet systems around M dwarfs and 9 planets in multi-planet systems around FGK dwarfs. This indicates that nearly 50\% of the M dwarf candidates are part of a systems of detectable planets, which is consistent with the compact multiple planet occurrence rate of 44\% from \citet{har19}. This could be due to the fact that smaller planets are more easily detected around M dwarfs, increasing the detectable planet multiplicity. With such a small sample (75 candidates) it remains difficult to make strong claims about the population parameters of M dwarf vs. FGK dwarfs planets, but overall they both follow similar period and radius distributions. A more thorough investigation of population parameters will be left for a future study.    

\subsection{Planet Radius}
\label{sec:plRad}
The \emph{K2} data set must undergo significant processing (Section \ref{sec:data}) before the transit may be fit. This extensive manipulation may lead one to question the accuracy of the extracted parameters. The feature that experiences the largest risk of modification is the transit depth. The {\tt EVEREST} software, the GP detrenders, and the harmonic fitter all have the potential to remove some amount of the transit depth. Fortunately, injecting signals at the light-curve-level (as done in Section \ref{sec:injections}) affords us the opportunity to measure the transit depth before and after processing.

\begin{figure}
\centering \includegraphics[width=\columnwidth]{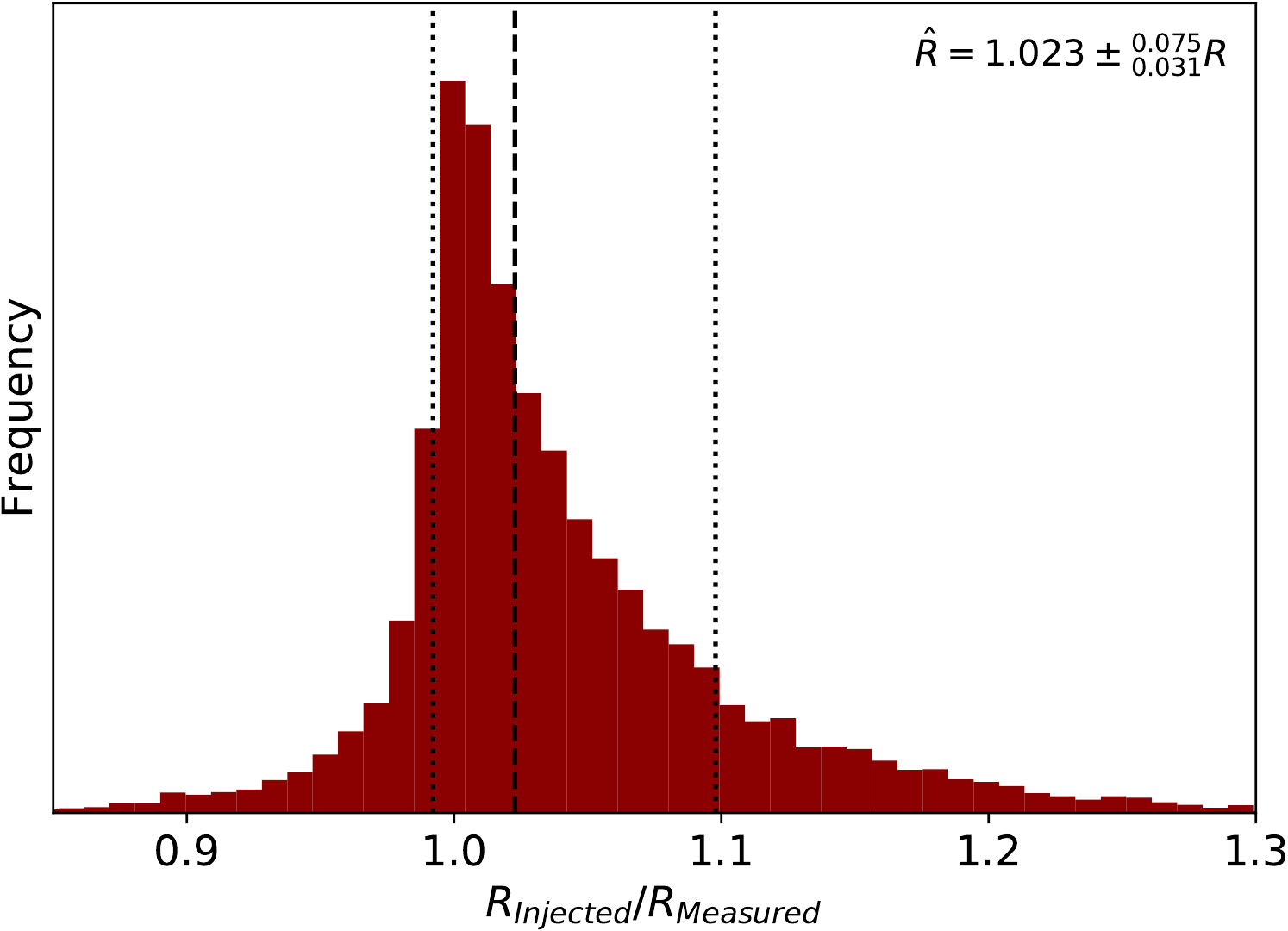}

\caption{A histogram of the injected to recovered planet radius ratio with bin widths of 0.01. The detrending and processing of the light curves on average leads to a $2.3\%$ reduction in the inferred planet radius. The dashed line represents the median ratio value and the dotted lines mark the 68\% quartiles. This shows that the offset has a statistical significance of $<1\sigma$.   \label{fig:radBias}}
\end{figure}

We focus on the planet radius, as this parameter has the largest impact on the transit depth. In Figure \ref{fig:radBias} we present the ratio of expected radius and recovered radius after processing. We find an expected offset of $2.3\%$ for the planet radius. This means, on average, our pipeline will underestimate the planet radius by $2.3\%$. In comparison, the \emph{Kepler} pipeline considered the offset in measured MES and detected MES (0.4\%), transforming this metric to radius, the expected offset for \emph{Kepler} is $-0.19\%$ for the recovered planets \citep{chr13}. Unsurprisingly, the \emph{K2} pipeline is far more prone to transit depth reduction. However, both of these offsets (\emph{Kepler} and \emph{K2}) are smaller than the average planet radius uncertainty ($\sim10\%$; \citealt{ber18}). Knowing that this offset falls within the normal radius uncertainty, Figure \ref{fig:radBias} shows the mode of this offset distribution is located at 1, we do not apply this correction to our candidate parameters. We leave such a procedure to be carried out at the reader's discretion. However, we do incorporate this potential offset into our measure of radius uncertainty. To account for the uncertainty expected by the radius offset ($\sigma_{\textrm{Off}}$), we use the following operation:

\begin{equation}
\sigma_{\textrm{Off}}=0.023\times R
\end{equation}
where $R$ is our best estimation of the planet radius. 

We also investigated the possible difference in this ratio distribution when separating long period injections ($>10$ days) from short period injections ($<10$ days). We expect planets with more available transits might be less prone to this type of depth reduction, but we find no significant difference between these two distributions. Therefore, we only focus on the combined population.

Beyond the radius offset mentioned above, we also want to consider our overestimation of flux contamination (Section \ref{sec:blend}). We assumed a 1D model for flux contamination, but a 2D model would produce less contamination ($\sim1.88\times$ in the worst case scenario). To address this issue, we fold the potential overestimation ($\sigma_F$) into our measure of radius uncertainty:

\begin{equation}
\begin{aligned}[t]
R & =R_{fit}\times R_{\star}\times \sqrt{F^{Tot}}\\
\left| \frac{\partial R}{\partial F^{Tot}} \right| & =\frac{R}{2\times F^{Tot}}\\
d F^{Tot} & =\frac{F^{Tot}-1}{1.88}\\
\sigma_F  = \left| \frac{\partial R}{\partial F^{Tot}} \right| d F^{Tot}  & = R\frac{F^{Tot}-1}{3.76\times F^{Tot}}\\
\end{aligned}
\end{equation}
where $R_{fit}$ is the radius ratio found using our MCMC, $R_{\star}$ is the radius of the stellar host, and $F^{Tot}$ is the flux ratio from the total flux to the flux of the target star. If $F^{Tot}$ is one, $\sigma_F$ is zero because no contamination was found, and therefore no correction is warranted. The uncertainty contribution from this parameter is very small ($\sigma_F/R\sim6.9\times10^{-6}$). Our combined estimation of planet radius uncertainty is as follows:

\begin{equation}
\sigma_R=\sqrt{\sigma_{fit}^2+\sigma_{\star}^2+\sigma_F^2+\sigma_{\textrm{Off}}^2}
\end{equation}
where $\sigma_{fit}$ is the uncertainty measured by our MCMC for $R_{fit}$ ($\sigma_{fit}/R\sim0.040$) and $\sigma_{\star}$ is the uncertainty measured for the stellar radius ($\sigma_{\star}/R\sim0.06$). In almost all cases $\sigma_{fit}$ and $\sigma_{\star}$ dominate the uncertainty, but the other parameters provide a slight increase to account for the issues they represent. Overall, we find a median$(\sigma_{R}/R)$ of 0.16 for our list of planet candidates.

The radius distribution recovered by our pipeline shows some hints of the radius gap near $2R_{\oplus}$ \citep{ful17}, shown by the double peak in the blue radius density plot of Figure \ref{fig:cand}. The improved stellar parameters available in this study allow this gap to be visible even with our small sample size of 75 candidates. Evidence for this gap in the \emph{K2} data has also been acknowledged in \citet{may18,kru19}. However, a more detailed accounting of completeness is necessary to verify such claims. We leave this task for future studies.

\section{Conclusions} \label{sec:conclusions}

We have provided the results from our pilot study, in which we created a uniform \emph{K2} Campaign 5 planet candidate catalog. The results of this study are intended to help generate a population of planets that can be used to measure the underlying exoplanet population for this campaign out to 38 days. We also show that our methodology can be applied to the remaining \emph{K2} C11-C18 campaigns, where the number of known candidates is much lower and cannot provide sufficient data for tuning of the pipeline. 

Our candidate sample includes 75 planets with seven multi-planet systems (5 double, 1 triple, and 1 quadruple planet systems), and nine new candidates. We have successfully recovered 51\% of the known planet candidates \citep{kru19} within the range of our survey, and can uniformly measure the completeness and reliability of our sample.

We also introduce our vetting software ({\tt EDI-Vetter}), which has been optimized to address issues specific to the \emph{K2} data set. This software builds upon the metrics of {\tt Robovetter} \citep{tho18}, which was used on \emph{Kepler} DR25. We provide several new features that allow us to produce a reliable candidate catalog for \emph{K2}. The philosophy of {\tt EDI-Vetter} is to minimize the number of FPs, even if it requires the loss of some true planet candidates, which is more difficult in \emph{K2} than \emph{Kepler} because of the systematic noise introduced by the spacecraft roll and thrusters.

To measure how reliable our pipeline is, we invert the light curves to ensure no true transits exist. The pipeline is then run, and any TCEs found and vetted can be considered true FPs. Of 6,145 TCEs found in the light curves, only four were able to deceive {\tt EDI-Vetter}. Using Equation \ref{eq:reli}, we found {\tt EDI-Vetter} to be 94.2\% reliable for our catalog. This means that statistically we should only expect four planets within our catalog to be systematic FPs. However, such tests will only address the possibility of systematic FPs. Astrophysical FPs, such as eclipsing binaries, can dilute the transit depth and lead to signal mis-classification \citep{ful17,mat18}. Additionally, binary stars can cause inaccurate planet radius measurements. \citet{cia17} showed that stellar multiplicity can lead to and overestimation in the number of Earth-sized planets by 15-20\%.

We also measured the completeness of our sample, and find a maximum completeness of 70\% for TCEs, and 60\% for vetted candidates. By injecting artificial transits at the light-curve-level of the raw \emph{K2} data, we find that our vetting software is only reducing the recovery rate by about 10\%. This is comparable to the loss introduced by {\tt Robovetter} (see Figure \ref{fig:kepk2}), indicating that our vetting software is not contributing to the majority of the recovery loss. The lack of completeness is in large part due to the systematic issues of \emph{K2}, increasing the noise and the likelihood of a transit being artificially removed (via detrending) from the light curve. 

To ensure occurrence studies can be carried out with this catalog, we provide CDPP and window function measurements for each stellar target. These data are available in a machine-readable version online.

\subsection{Recommendations for Occurrence Studies}
\label{sec:rec}
For occurrence studies we recommend the following procedures:
\begin{itemize}
  \item Use a stellar sample where all light curves have a CDPP measure $\le1,200$ ppm for a transit window of 8 hours. Light curves with more noise are problematic and unlikely to yield usable transits.
  \item Only use a target light curve if the window function has a probability $>0.8$ for periods of 18 days. Targets with lower probabilities are missing significant portions of the light curve, making transit detections difficult.
  \item Consider the issue of systematic correction for radius as seen in Figure \ref{fig:radBias}. This effect could alter the inferred radius distribution of this population of planets. 
    \item Make an upper limit cut on the planet radius populations. Our eclipsing binary metric is agnostic to stellar radius in order to eliminate any potential biases a specific stellar catalog may introduce. Thus, large radius planet candidates ($R>20R_\Earth$) may be astrophysical false positives that slipped through our vetting metrics (further discussion on this topic in Section \ref{sec:candidates}).
    \item The measure of completeness provided in this study assumes all planets meet the criteria presented in Equation \ref{eq:EB}. It is possible that real transiting planets that graze the limb of a star may not meet this criteria. To account for this possibility, we suggest future occurrence studies use the follow probability function to account for this issue:
    \begin{equation}
    \resizebox{.85 \columnwidth}{!}{$
P\bigg(\text{Detection}\bigg|\frac{R_{pl}}{R_{\star}}\bigg)=$ $
\begin{cases}
           1; & \frac{R_{pl}}{R_{\star}}\le0.04\\
           
           1.04-R_{pl}/R_{\star}; &  0.04<\frac{R_{pl}}{R_{\star}}\le0.3 \\
           
           0; & 0.3<\frac{R_{pl}}{R_{\star}}\\ 
    \end{cases}
    $}
    \end{equation}
    
    where $P$ is the probability of detection, assuming a uniform distribution of impact parameters from 0 to 1. This correction is minimal for small measures of the planet to star radius ratio ($R_{pl}/R_{\star}$). However, in cases where $R_{pl}/R_{\star}$ is larger, usually in planets orbiting M dwarfs, this probability can get as low as 74\%.       
    
\end{itemize}

\section{Acknowledgements} 
 We immensely thank the anonymous referee for the careful reading and very thorough comments.

Additionally, we would like to thank Christina Hedges, Geert Barentsen, and Jessie Dotson at the \emph{Kepler} Guest Observer office for helpful discussions about \emph{K2} and this work.

Trevor David provided helpful discussion of K2 light curve detrending and transit injections.

An additional special thanks to Megan Bedell for providing us with Gaia-\emph{Kepler} crossmatching out to radius $20\arcsec$ upon special request. Also we would like to thank Melanie Swain from the \href{https://exofop.ipac.caltech.edu}{ExoFOP} team for helping provide a massive export of stellar data upon request.

This work made use of the gaia-kepler.fun crossmatch database created by Megan Bedell. The simulations described here were performed on the UCLA Hoffman2 shared computing cluster and using the resources provided by the Bhaumik Institute. This research has made use of the NASA Exoplanet Archive and the Exoplanet Follow-up Observation Program website , which are operated by the California Institute of Technology, under contract with the National Aeronautics and Space Administration under the Exoplanet Exploration Program. This paper includes data collected by the \emph{Kepler} mission and obtained from the MAST data archive at the Space Telescope Science Institute (STScI). Funding for the \emph{Kepler} mission is provided by the NASA Science Mission Directorate. STScI is operated by the Association of Universities for Research in Astronomy, Inc., under NASA contract NAS 5–26555.

Guoshoujing Telescope (the Large Sky Area Multi-Object Fiber Spectroscopic Telescope LAMOST) is a National Major Scientific Project built by the Chinese Academy of Sciences. Funding for the project has been provided by the National Development and Reform Commission. LAMOST is operated and managed by the National Astronomical Observatories, Chinese Academy of Sciences.

Funding for this project was provided by the IPAC Visiting Graduate Fellowship. J. Z. acknowledges funding from NASA ADAP grant 80NSSC18K0431.

\software{{\tt EVEREST} \citep{lug16,lug18}, {\tt TERRA} \citep{pet13b}, {\tt K2SFF} \citep{van14}, {\tt K2PHOT} \citep{pet13b}, {\tt PyMC3} \citep{sal15}, {\tt Exoplanet} \citep{for19}, {\tt RoboVetter} \citep{tho18}, {\tt batman} \cite{kre15}, {\tt emcee} \citep{for13}}

\bibliography{bib.bib}\setlength{\itemsep}{-2mm}



\end{document}